%% file: main.tex
\documentclass[twocolumn]{aastex62}

\usepackage[utf8]{inputenc}
\usepackage{amsfonts}
\usepackage{amsmath}
\usepackage{mathrsfs}
\usepackage{amssymb}
\usepackage{textgreek}
\usepackage[nolist,nohyperlinks,printonlyused]{acronym}
\usepackage[english]{babel}
\usepackage{savesym}
\usepackage{verbatim}
\usepackage[bottom]{footmisc}

\usepackage [autostyle, english = american]{csquotes}
\MakeOuterQuote{"}

\renewcommand\plotone[1]{%
 \centering
 \leavevmode
 \includegraphics[width={0.95\linewidth}]{#1}%
}%
\usepackage{printlen}
\savesymbol{tablenum}
\usepackage{siunitx}
\restoresymbol{SIX}{tablenum}

\bibpunct{(}{)}{;}{a}{}{,}

\newcommand{\vect}[1]{\boldsymbol{#1}}

\newcommand{\HI}{\ion{H}{1}}
\newcommand{\HII}{\ion{H}{2}}
\newcommand{\HeI}{\ion{He}{1}}
\newcommand{\HeII}{\ion{He}{2}}

\newcommand{\SiIII}{\ion{Si}{3}}

\newcommand{\mathHI}{{\mbox{\scriptsize \HI}}}
\newcommand{\mathHII}{{\mbox{\scriptsize \HII}}}

\newcommand{\mathHeII}{{\mbox{\scriptsize \HeII}}}

\newcommand{\mathSiIII}{{\mbox{\scriptsize \SiIII}}}

\DeclareSIUnit \parsec{pc}
\DeclareSIUnit \h {\ensuremath{\mathit{h}}}
\DeclareSIUnit\com{com. }
\DeclareSIUnit\yr{yr}

\newcommand{\skm}{\second\per\kilo\meter}
\newcommand{\Mpc}{\mega\parsec}
\newcommand{\kpc}{\kilo\parsec}
\newcommand{\hinvMpc}{\per\h\Mpc}

\newcommand{\lyatitle}{L\MakeLowercase{yα}}

\def\equationautorefname~#1\null{eqn.~(#1)\null}
\submitjournal{ApJ}
\accepted{12/20/2018}

\begin{document}
\title{New Constraints on IGM Thermal Evolution from the \lyatitle{} Forest Power Spectrum}

\author[0000-0001-9182-6972]{Michael Walther}
\correspondingauthor{Michael Walther}
\affiliation{Physics Department, Broida Hall, University of California Santa Barbara, Santa Barbara, CA
93106-9530, USA}
\affiliation{Max-Planck-Institut für Astronomie, Königstuhl 17,
69117 Heidelberg, Germany}
\affiliation{International Max Planck Research School for Astronomy \&
Cosmic Physics at the University of Heidelberg}

\author[0000-0002-8723-1180]{Jose O\~norbe}
\affiliation{Royal Observatories, Blackford Hill, Edinburgh EH9 3HJ, UK}
\affiliation{Max-Planck-Institut für Astronomie, Königstuhl 17,
69117 Heidelberg, Germany}

\author[0000-0002-7054-4332]{Joseph F. Hennawi}
\affiliation{Physics Department, Broida Hall, University of California Santa Barbara, Santa Barbara, CA
93106-9530, USA}
\affiliation{Max-Planck-Institut für Astronomie, Königstuhl 17,
69117 Heidelberg, Germany}

\author{Zarija Luki\'c}
\affiliation{Lawrence Berkeley National Laboratory, 1 Cyclotron Road, Berkeley, CA
94720, USA}

\email{\bf michael.walther@cea.fr}

\defcitealias{Walther2017NewPrecision}{\textsc{Paper I}}
\defcitealias{palanque-delabrouille2013onedimensionalLy}{\textsc{PD+13}}

\begin{abstract}
We determine the thermal evolution of the intergalactic medium (IGM)
over \(\SI{3}{\giga\yr}\) of cosmic time \(1.8<z<5.4\) by comparing
measurements of the Ly\(\alpha\) forest power spectrum to a suite of
\(\sim70\) hydrodynamical simulations.  We conduct Bayesian inference
of IGM thermal parameters using an end-to-end forward modeling
framework whereby mock spectra generated from our simulation grid are
used to build a custom emulator which interpolates the power spectrum
between thermal grid points.  The temperature at mean density $T_0$
rises steadily from \(T_0\sim\SI{6000}{K}\) at \(z=5.4\), peaks at
\(\SI{14000}{K}\) for \(z\sim 3.4\), and decreases at lower redshift
reaching \(T_0\sim\SI{7000}{K}\) by \(z\sim1.8\).  This evolution
provides conclusive evidence for photoionization heating resulting
from the reionization of \HeII{}, as well as the subsequent cooling
of the IGM due to the expansion of the Universe after all reionization
events are complete. Our results are broadly consistent with previous
measurements of thermal evolution based on a variety of approaches, but the
sensitivity of the power spectrum, the combination of high precision
BOSS measurements of large-scale modes (\(k\lesssim\SI{0.02}{\skm}\))
with our recent determination of the small-scale power, our large grid
of models, and our careful statistical analysis allow us to break the
well known degeneracy between the temperature at mean density \(T_0\)
and the slope of the temperature density relation \(\gamma\) that has
plagued previous analyses.  At the highest redshifts \(z\geq5\) we
infer lower temperatures than expected from the standard picture of IGM thermal
evolution leaving little room for additional smoothing of the Ly\(\alpha\) forest by
free streaming of warm dark matter.

\end{abstract}

\section{Introduction}

The \ac{Lya} forest \citep{gunn1965DensityNeutralHydrogen,Lynds1971Absorption-LineSpectrum} is the premier probe of diffuse baryons in the \ac{IGM} at high redshifts.
Its fluctuations can be accurately described in the current \(\Lambda \mathrm{CDM}\) framework --- on large scales it is mostly sensitive to cosmological parameters such as the amplitude of fluctuations \(\sigma_8\), primordial power spectrum slope \(n_s\), baryon density \(\Omega_b\), number of neutrino species \(N_\mathrm{eff}\), and the sum of neutrino masses \(\sum m_\nu\) \citep{Palanque-Delabrouille2015Neutrinomasses,Rossi2017ImpactMassive}.
On small scales, however, it is sensitive to the thermal state of the \ac{IGM}\footnote{Note that the small scale \ac{Lya} forest is also sensitive to the nature of dark matter \citep[like \ac{WDM}, see e.g.][]{Seljak2006CanSterile,viel2013Warmdarkmatter} which will not be the focus of this work, but leads to important applications of our results.}.
This  alters the observed spectra via the Doppler broadening of absorption features due to thermal motions, as well as pressure smoothing of the gas (sometimes called "Jeans" broadening), which affects the underlying baryon distribution and depends on the integrated thermal history of the \ac{IGM} \citep{gnedin1998ProbingUniverseLyalpha, kulkarni2015CharacterizingPressureSmoothing,Onorbe2017Self-consistentModeling}. The thermal evolution is largely driven by impulsive heating from cosmic reionization events and the cooling process due to adiabatic expansion and Compton cooling \citep{McQuinn2016intergalactictemperature-density}.

Current constraints imply that hydrogen and \HeI{} were reionized at \(z_\mathrm{reion, 50}= 6.4\textrm{--}9.0 (95\%)\)\footnote{\(z_\mathrm{reion, 50}\) is the redshift at which \(x_\mathHI= 0.50\).} (see \citealt{PlanckCollaboration2018Planck2018}).
Additionally, measurements of the \ac{Lya} forest optical depth show a strong
increase close to \(z=6\) leading to complete Gunn-Peterson absorption
\citep{fan2006Constrainingevolutionionizing,
  becker2015Evidencepatchyhydrogen, Bosman2018Newconstraints,
  Eilers2018OpacityIntergalactic} which reveals that reionization ends at
\(z\sim 6\).  As for \HeII{} reionization, which is driven by the
hard \(>4\) Ryd photons emitted by luminous quasars,  observations of the \HeII{}
Ly\(\alpha\) forest indicate \HeII{} had to be reionized by \(z=2.7\)
\citep{worseck2011EndHeliumReionization} and possibly
as early as  \(z=3.4\) \citep{Worseck2016EarlyExtended}, but the limited
number of observational constraints imply that the exact timing
remains largely uncertain.
While it is observationally tricky to obtain direct higher redshift constraints on \HeII{} reionization through \HeII{} \ac{Lya} absorption measurements because the \HeII{} forest becomes more and more opaque, we can indirectly constrain it via its imprint on the thermal state of the IGM.

In the standard picture of thermal evolution cold \ac{IGM} gas (few \(\si{\kelvin}\)) is strongly heated during \HI{} and \HeI{} reionization (by few times \(\SI{1e4}{\kelvin}\)), subsequently cools and then experiences additional heating during \HeII{} reionization \citep{mcquinn2009HeIIReionization,Compostella2013imprintinhomogeneous,puchwein2015photoheatingintergalacticmedium,Greig2015impacttemperature,uptonsanderbeck2016Modelsthermalevolution,McQuinn2016intergalactictemperature-density,Onorbe2017Self-consistentModeling,Puchwein2018Consistentmodelling}.
The combined effects of photoionization heating,  Compton cooling, and adiabatic cooling due to the expansion of the universe lead to a net cooling of intergalactic gas between and after the reionization phases which has so far not been conclusively observed.
Another consequence of these effects is a tight power law \ac{TDR} for most of the \ac{IGM} gas \citep{hui1997Equationstatephotoionized, puchwein2015photoheatingintergalacticmedium, McQuinn2016intergalactictemperature-density}
about \(\Delta z\approx 1\text{--}2\) after the impulsive heating from
a reionization event:
\begin{equation}
\label{eq:t-rho-relation}
 T(\Delta) = T_0 \Delta^{\gamma-1},
\end{equation}
where \(\Delta = \rho / \bar{\rho}\) is the overdensity, \(T_0\) is temperature at mean density \(T_0\), and the
index \(\gamma\) is expected to approach \(\sim 1.6\) long after the completion of reionization.

As we recently summarized in \citet{Walther2017NewPrecision} (hereafter \citetalias{Walther2017NewPrecision}) there have been many attempts to measure the IGM's thermal parameters \citep{haehnelt1998Probingthermalhistory,schaye2000thermalhistoryintergalactic,bryan2000DistributionLyForest,ricotti2000EvolutionEffectiveEquation,mcdonald2001MeasurementTemperatureDensity,theuns2002Temperaturefluctuationsintergalactic,bolton2008Possibleevidenceinverted,Viel2009Cosmologicalastrophysical,lidz2010MeasurementSmallscale,becker2011DetectionextendedHe,rudie2012TemperatureDensityRelation,garzilli2012intergalacticmediumthermal,rorai2013NewMethodDirectly,viel2013Warmdarkmatter, boera2014thermalhistoryintergalactic,bolton2014consistentdeterminationtemperature,Lee2015IGMConstraints,Rorai2017Exploringthermal,Rorai2017Measurementsmall-scale,Irsic2017Newconstraints,yeche2017Constraintsneutrinomasses,Garzilli2017CutoffLyman-a,Rorai2018newmeasurement,DAloisio2018Largefluctuations,Hiss2017NewMeasurement} based on different statistical techniques which typically constrain the smoothness of the \ac{Lya} forest as a whole via some summary statistics (e.g. wavelet amplitudes, spectral curvature or the power spectrum) or decompose the forest into individual absorption lines by Voigt profile fitting.
While there were some notable discrepancies between some of the older measurements (e.g. low values of
\(\gamma\) inferred from the \citealt{bolton2008Possibleevidenceinverted} flux PDF or the high \(T_0\) measurements
from the \citealt{lidz2010MeasurementSmallscale} wavelet analysis), more recent measurements appear to be in better agreement. For example,
temperature determinations from the curvature statistic \citep{becker2011DetectionextendedHe,boera2014thermalhistoryintergalactic} agree fairly well with those determined from Voigt profile fitting \citep{bolton2014consistentdeterminationtemperature,Rorai2018newmeasurement,Hiss2017NewMeasurement}.
Note however, that different techniques have distinct systematics and parameter degeneracies, that often complicate
detailed comparisons.

In this work, we use the power spectrum of the \ac{Lya} forest to obtain an accurate self-consistent measurement of IGM thermal evolution  over a large redshift range from \(z=5.4\) to \(z=1.8\).
The power spectrum exhibits a cutoff at small scales (high \(k\)) beyond which there is no structure left in the \ac{Lya} forest.
The reason for this is both the smoothness in the baryon density resulting from the finite gas pressure (often called Jeans pressure smoothing) as well as thermal Doppler broadening.
The great advantage of the power spectrum compared to other methods, is its sensitivity to structure on a multitude of scales.
Specifically, whereas other methods like the curvature \citep{becker2011DetectionextendedHe} and wavelets \citep{lidz2010MeasurementSmallscale} provide only a small-scale measurement of
spectral smoothness, the overall shape of the power spectrum for scales between \(\sim \SI{500}{\kilo\parsec}\) and \(\sim \SI{10}{\mega\parsec}\) as well as small-scale (high-k) cutoff provides additional constraining power that breaks degeneracies between different thermal parameters\footnote{Note that this property can also be used to break degeneracies with cosmological parameters, e.g. the nature of dark matter  \citep{viel2013Warmdarkmatter, Irsic2017Newconstraints, armengaud2017Constrainingmasslight} or the mass of neutrinos \citep{Palanque-Delabrouille2015Neutrinomasses, yeche2017Constraintsneutrinomasses, baur2017ConstraintsLybackslash}.}.
For this work we consider \(T_0\), \(\gamma\) and the pressure smoothing scale \(\lambda_P\) as thermal parameters and the mean transmission \(\bar{F}\) as a further astrophysical parameter. We additionally marginalize over the strength of \SiIII{} correlations and the resolution of the X-SHOOTER spectrograph (see \autoref{sec:priors} for more detailed information about our prior assumptions).

Our analysis is based upon our recent high-precision measurements of the the small-scale (high wavenumber \(k\)) the \ac{Lya} forest flux power spectrum in \citetalias{Walther2017NewPrecision} as well as other recent measurements from different instruments (\citealt{palanque-delabrouille2013onedimensionalLy}, hereafter \citetalias{palanque-delabrouille2013onedimensionalLy}; \citealt{viel2013Warmdarkmatter}; and \citealt{Irsic2017Newconstraints}) combined with the new \ac{THERMAL} grid\footnote{see \url{thermal.joseonorbe.com}} of hydrodynamical simulations.
We then perform inference by employing fast interpolation of our model power spectra and performing an MCMC analysis with a Gaussian likelihood.

This paper is organized as follows. The measurements we used in this work are summarized in \autoref{sec:measurements}.
In \autoref{sec:sims} we present our grid of hydrodynamical simulations.
We use modified versions of our forward modeling, interpolation and inference tools from \citetalias{Walther2017NewPrecision}, which we present in \autoref{sec:method}, to measure the thermal state of the IGM at each redshift.
In \autoref{sec:results} we present these results and compare them to measurements from the literature as well as thermal evolution models.
Finally, we discuss the results in \autoref{sec:discussion} and conclude with \autoref{sec:conclusions}.

\section{Power Spectrum Datasets for Studying IGM Thermal Evolution}
\label{sec:measurements}
\input{datatable}
In \citetalias{Walther2017NewPrecision} we performed a new measurement of the \ac{Lya} forest power spectrum.
This is based on 74 archival high-resolution, high-S/N quasar spectra obtained with the VLT/UVES \citep[from][]{dallaglio2008unbiasedmeasurementUV} and Keck/HIRES \citep[from][]{omeara2015FirstDataReleasea,omeara2017kodiaqdr2} spectrographs covering a redshift range from \(z=1.8\) to \(z=3.4\).
This comprises a significant improvement in dataset size compared to previous measurements based on high-resolution spectra \citep{mcdonald2000ObservedProbabilityDistribution,croft2002PreciseMeasurementMatter,kim2004powerspectrumflux,viel2008HowColdCold} in this redshift range.
We semi-automatically masked out possible metal contamination in our data based on several approaches, measured the power spectrum using a Lomb-Scargle Periodogram \citep{lomb1976Leastsquaresfrequency,scargle1982Studiesastronomicaltime} on the flux contrast \(\delta_F=(F-\bar{F})/\bar{F}\), and binned the resulting power in equidistant bins in \(\log k\).
Statistical uncertainties were estimated using a bootstrap method and are \(\lesssim 10\%\)
for the small scale modes that are most sensitive to the thermal state of the \ac{IGM}.

Additionally, data using the BOSS \citep[Baryon Oscillation Spectroscopic Survey, with the dataset of ][]{palanque-delabrouille2013onedimensionalLy} or X-SHOOTER \citep[datasets of ][]{Irsic2017Lymana,yeche2017Constraintsneutrinomasses} spectrographs are available with even smaller statistical uncertainties (e.g. \(\sim 2\%\) on large scales \(k<\SI{0.01}{\skm}\) for the BOSS dataset),
but limited small scale power spectrum coverage due to the significantly lower spectroscopic resolutions of these instruments.
As these analyses use the same redshift binning as we do, but extend to higher redshifts  \(3.6 \leq z \leq 4.2\) a comparison to them is straightforward.
In particular, the BOSS data provides a large scale anchor point thereby partially breaking degeneracies between the different parameters.
However, the X-SHOOTER dataset may have significant uncertainty in its resolution estimates which we will take into account in our modeling procedure (see \autoref{sec:priors})\footnote{This issue was discussed in \citetalias{Walther2017NewPrecision}. See also \citet{Selsing2018X-shooterGRB} who show the dependence of spectroscopic resolution on seeing for the VIS and NIR arms in their Fig. 2 and find both significant scatter as well as overall higher resolution than previously quoted on the ESO webpage.}.

To assess the thermal state at even higher redshifts \(4.2\leq z\leq 5.4\) (where currently no large survey dataset exists) we use data from the previous high-resolution measurement by \citet{viel2013Warmdarkmatter} based on Keck/HIRES and Magellan/MIKE data.
This extension allows us to cover a big part of the universes history (\(1.8<z<5.4\)) from just  after \HI{} reionization to well after the \HeII{} reionization \citep[according to][]{Worseck2016EarlyExtended} and the peak of the cosmic star formation history.

To summarize, our fiducial dataset consists of the data from \citetalias{Walther2017NewPrecision} for \(z\leq 3.4\), the BOSS data by \citetalias{palanque-delabrouille2013onedimensionalLy} at \(2.2\leq z\leq4.2\), the data by \citet{viel2013Warmdarkmatter} at \(z\geq 4.2\), and the XQ-100 measurement by \citet{Irsic2017Lymana} at \(3.6\leq z\leq 4.2\) where the VIS arm was used (for \(z=3.6\) jointly with data from the UVB arm). Note that for \(3.6\leq z\leq 4.0\) no recent high-resolution analysis is available. A summary of the datasets we analyzed can be found in \autoref{tab:datasets}. Here, we show the observed redshift range \(z_\mathrm{min}\)--\( z_\mathrm{max}\) , the binning in redshift \(\Delta z\), the number of spectra analyzed \(N_\mathrm{qso}\), the approximate resolution \(R\), and the maximal wavenumber \(k_\mathrm{max}\) obtained.

\section{The THERMAL Suite of Hydrodynamical Simulations}
\label{sec:sims}
\label{sec:hydro}

The hydrodynamical models we use in this paper for comparison with our measurement are part of the publically available \ac{THERMAL} suite of
Nyx simulations \citep{almgren2013NyxMassivelyParallel}.
Nyx follows  the  evolution  of  dark  matter simulated as self-gravitating Lagrangian particles, and baryons modeled as an ideal gas on a uniform Cartesian grid.
The Eulerian gas dynamics equations are solved using a second-order accurate \ac{PPM} to accurately capture shocks.
For more details of these numerical methods and scaling behavior tests, see \citet{almgren2013NyxMassivelyParallel} and \citet{lukic2015Lymanforestoptically}.

Besides solving for gravity and the Euler equations, we also include
the main physical processes fundamental to model the \ac{Lya}
forest. First we consider the chemistry of the gas as having a
primordial composition with hydrogen and helium mass abundances of
$X_\mathrm{p}$, and $Y_\mathrm{p}$, respectively.  In addition, we include
inverse Compton cooling off the microwave background and keep track of
the net loss of thermal energy resulting from atomic collisional
processes.  We used the updated recombination, collision ionization,
dielectric recombination rates, and cooling rates given in
\citet{lukic2015Lymanforestoptically}.  All cells are assumed to be
optically thin to ionizing radiation, and radiative feedback is
accounted for via a spatially uniform, but time-varying \ac{UVB}
radiation field given to the code as a list of photoionization and
photoheating rates that vary with redshift
\citep[e.g.][]{katz1992Galaxiesgascold}.

The THERMAL suite consists of \(\sim 70\) simulations, each in \(L_\mathrm{box}=\SI{20}{\hinvMpc}\) box and using \(N_{cell}=1024^3\) Eulerian cells and \(1024^3\) dark matter particles which is a strong improvement with respect to previous studies of the thermal state which relied on smaller boxes with the same resolution \citep[e.g.][]{becker2011DetectionextendedHe}.
Cosmology is based on a \citet{planckcollaboration2014Planck2013results} model
(\(\Omega_m=0.319181\), \(\Omega_b h^2=0.022312\), \(h=0.670386\),
\(n_s=0.96\), \(\sigma_8=0.8288\)). Comparisons of different
resolutions and box sizes can be found in
\citet{lukic2015Lymanforestoptically} and this box size was chosen as
the best compromise between being able to run a large grid of models
and the need to be converged at least to \(< 10\%\) on small scales
(large \(k\)).
The power spectrum is even converged to the one percent level on all relevant scales for \(z\lesssim 3\) and all scales \(k\lesssim \SI{0.05}{\skm}\) at higher redshifts with respect to resolution.
For boxsize, however, the power is converged to the \(\sim 5\%\) level, with the largest scales (smallest \(k<\SI{0.01}{\skm}\)) being significantly influenced by poor mode sampling and therefore excluded from our analysis.
We further discuss effects of numerical convergence in \autoref{sec:discussion} which proves to be a major systematic effect for our analysis.

For most simulations we generated different thermal histories in a similar way as in \citet{becker2011DetectionextendedHe} by changing the heating rates relative to a fiducial model at all redshifts and we'll henceforth call these our `heating rate rescaling models'.
The heating rates we used to construct different thermal histories have been constructed as:
\begin{equation}
\label{eq:heating-rates}
  \epsilon=A \Delta^B \epsilon_\textup{HM12},
\end{equation}
where \(\epsilon_\textup{HM12}\) are the heating rates tabulated in \citet{haardt2012RadiativeTransferClumpy} and \(A\) and \(B\) are the parameters changed to get different thermal histories.
Note that while long after any reionization event the instantaneous temperature is more or less independent of the redshift of reionization, the pressure smoothing scale \(\lambda_\mathrm{P}\) retains a memory of this for a longer time (\citealt{gnedin2003LinearGasDynamics,kulkarni2015CharacterizingPressureSmoothing,Onorbe2017Self-consistentModeling}, an alternative parametrization is possible using the total heat input, see \citealt{Nasir2016InferringIGM}).
As this type of modeling leads to changes in the thermal state at all redshifts, it is hard to disentangle \(\lambda_\mathrm{P}\) from \(T_0\) and \(\gamma\) from just this approach.

Because of this and to better explore the parameter space we also use a second modeling approach providing completely distinct thermal histories.
In this approach we self-consistently solve for the UV background as well as the heating during reionization following the approach laid out in \citet{Onorbe2017Self-consistentModeling}.  Reionization models are parametrized by both a total heat input \(\Delta T\) during reionization and a redshift of reionization \(z_\mathrm{reion}\) (at which a species is \(99.9\%\) ionized and assuming a fixed shape for the reionization history) for both \HI{} and \HeII{} reionization.
We also consider the thermal histories based on this approach to be more physically motivated and will later use them to study the implications of our measurements on reionization.

The values for thermal parameters \(T_0\) and \(\gamma\) were obtained from the simulation by fitting a power law \ac{TDR} to the distribution of gas cells in \(\log \Delta\) and \(\log T\) using a linear least squares method as described in \citet{lukic2015Lymanforestoptically}.
To determine the pressure smoothing scale \(\lambda_\mathrm{P}\) the cutoff in the power spectrum  of the real-space \ac{Lya} flux \(F_\mathrm{real}\) was fit as described in \citet{kulkarni2015CharacterizingPressureSmoothing}.
Here, \(F_\mathrm{real}\) is the flux each position in the simulation would produce (given it's temperature and density), but neglecting redshift space effects.

\begin{figure*}
\centering
\plotone{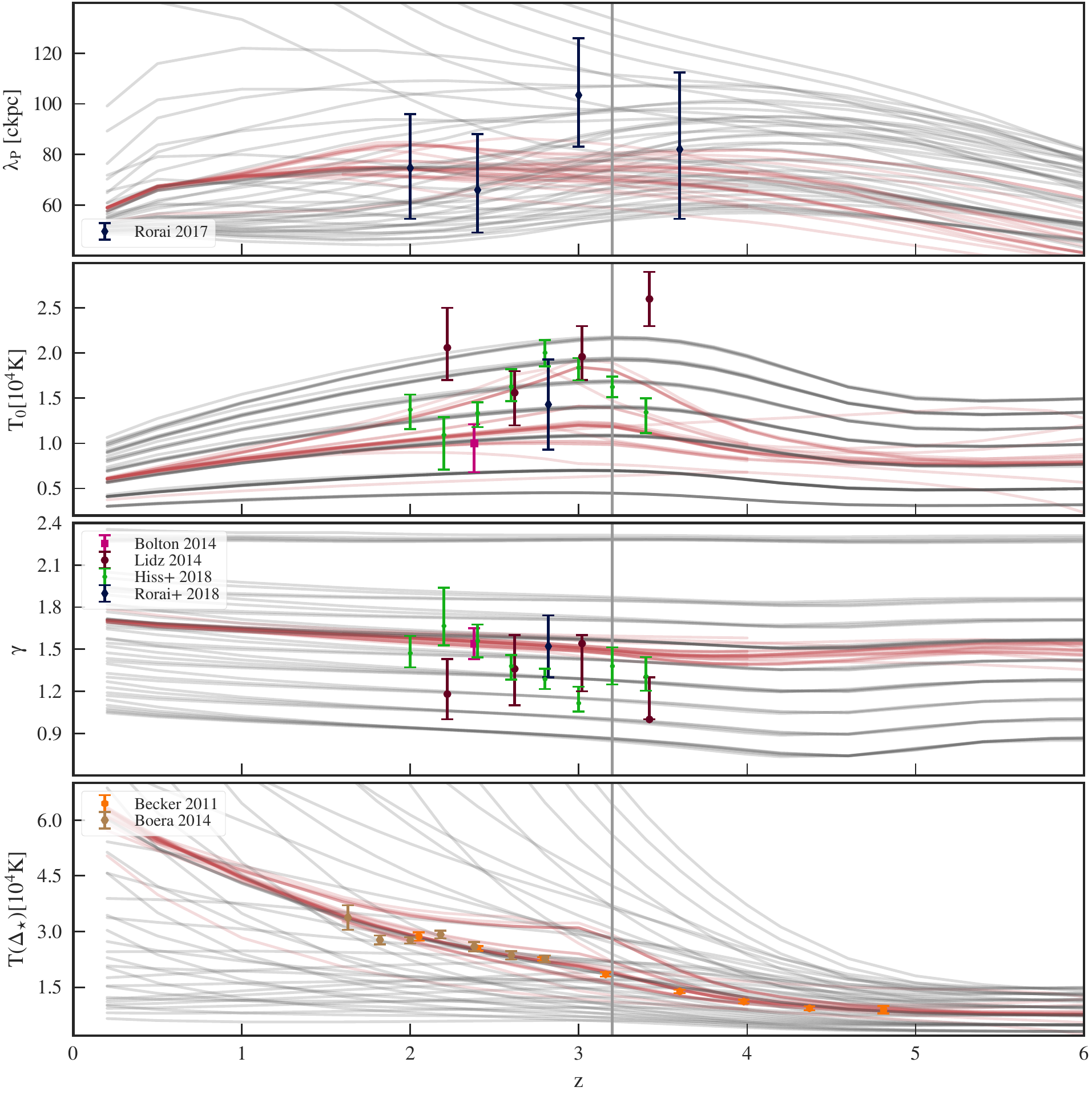}
\caption{Redshift evolution of different thermal evolution models (lines).
Most of the different curves (gray) where obtained by changing the overall heating rates from \citep{haardt2012RadiativeTransferClumpy} by a factor
(changing \(T_0\) at all redshifts) as well as the exponent in their density dependence (changing \(\gamma\) at all redshifts) according to \autoref{eq:heating-rates}.
As the pressure smoothing scale \(\lambda_\mathrm{P}\) is dependent on the full thermal evolution of the IGM it changes accordingly in these cases.
Additional models of thermal evolution (red) with different \HI{}
and \HeII{} reionization redshifts and heat inputs partially break those degeneracies.
Also shown is the Temperature \(T(\Delta_\star)\) \citep[based on the values of \(\Delta_\star\) by][]{becker2011DetectionextendedHe} at the overdensity where constraints from curvature measurements are independent of \(\gamma\).
We compare to the measurements by \citet{lidz2010MeasurementSmallscale}, \citet{becker2011DetectionextendedHe}, \citet{bolton2014consistentdeterminationtemperature}, \citet{boera2014thermalhistoryintergalactic}, \citet{Rorai2017Measurementsmall-scale}, \citet{Hiss2017NewMeasurement} and \citet{Rorai2018newmeasurement} in the parameters constrained by the respective analysis. The \citet{lidz2010MeasurementSmallscale}, \citet{bolton2014consistentdeterminationtemperature} and \citet{Rorai2018newmeasurement} data have been offset by 0.02 along the redshift axis for clarity.}\label{fig:redshift-evo-comp}
\end{figure*}

The model parameters were chosen to bracket most current observational constraints on thermal parameters from curvature, wavelet, line-fitting and quasar-pair phase angle statistics.
The set of all thermal evolution models used in this paper as well as the current observational constraints are shown in \autoref{fig:redshift-evo-comp}.
The explicit reionization based models (red curves) show strongly different evolutionary behavior especially  in \(T_0\) (most of them show a relatively narrow \HeII{} reionization peak around \(z=3\)) compared to a relatively smooth evolution for the heating rate rescaling approach (gray curves) and will also be used later as comparison models for our measured thermal evolution.

\begin{figure}
\plotone{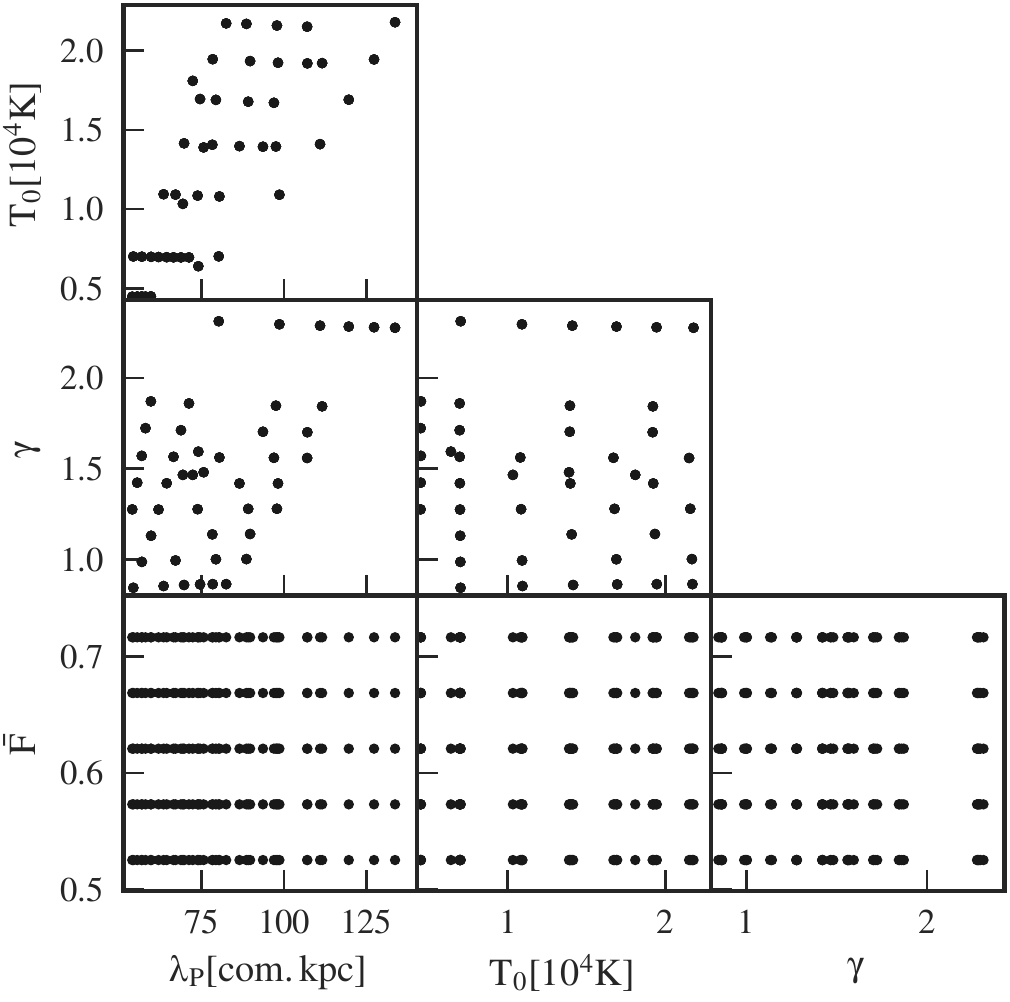}
\caption{The subset of thermal models we used at \(z=3.2\). Each point in the \(T_0,\gamma,\lambda_\mathrm{P}\) space corresponds to a different Hydrodynamical simulation.
Note the correlations between \(T_0\) (\(\gamma\)) and \(\lambda_\mathrm{P}\) in the grid.
For each simulation we rescaled the optical depths to obtain outputs with different mean transmitted fluxes \(\bar{F}\).}
\label{fig:grid_singlez}
\end{figure}

The combined set of models results in an irregular grid of thermal
parameters at each individual redshift. This is shown in
\autoref{fig:grid_singlez} where each point in the
\(T_0,\gamma,\lambda_\mathrm{P}\) volume corresponds to one of our
hydrodynamical simulations. We can see that a large range is spanned
in each of the parameters and most of the 2 parameter combinations.
As \(\lambda_\mathrm{P}\) probes the integrated thermal history which is smooth for each individual model and partly constrained by physical limits on heating and cooling of the IGM during and after reionization
it turns out to be relatively difficult to independently vary \(\lambda_\mathrm{P}\) in a way that is not correlated with the thermal state parameters \(T_0\) and \(\gamma\).
Alternatively, one could generate models with abruptly changing
temperature such that the pressure smoothing does not have enough time
to follow this change.  While arbitrary \(\lambda_\mathrm{P}\) could be
generated in this way, fine-tuning is needed to produce this kind of model
for an individual redshift which would take a lot of additional
computational time (especially for changes at low redshifts) and it also seems unphysical.
Therefore, we do not have full flexibility (mostly due to CPU time restrictions) in varying \(T_0\) vs. \(\lambda_\mathrm{P}\) orthogonal to the degeneracy direction visible in our models.
However, this in the end does not pose a problem to our analysis as the correlation between both parameters is physically motivated.

In principle reionization is an inhomogeneous process \citep{Davies2016Largefluctuations,DAloisio2015LARGEOPACITY}, but we only use an homogeneous model to describe photoionizations.
While generally \ac{UVB} and thermal fluctuations could be influencing the power spectrum and therefore our conclusions on thermal evolution especially at  \(z>4\) \citep[see e.g.][]{Cen2009ProbingEpoch}, recent analyses (\citealt{Onorbe2018Newmodels},  also earlier studies by \citealt{mcdonald2005PhysicaleffectsLy} and \citealt{Croft2004IonizingRadiation} obtained similar results but with a focus on lower redshifts) have found that those mostly change the power spectrum on larger scales than used for this work (at least for \HI{} reionization), but does not strongly change the power on small scales which provides most of the sensitivity to the thermal state of the \ac{IGM}.
Note again that we are not using the largest scale modes which strongly reduces our sensitivity to inhomogeneities, further justifying our use of a homogeneous \ac{UVB}.

We computed skewers of optical depth \(\tau=-\ln F\) by convolving each pixel along one dimension in the simulation box with the corresponding Voigt-profile for the Temperature \(T\), \(N_{\mathHI}\propto \Delta^2/(T^{0.7} \Gamma_{\mathHI})\) and Doppler shifts due to \(v\) for each simulation snapshot.
As is common in \ac{Lya} forest studies \citep[see e.g.][]{Bolton2010firstdirecta,boera2014thermalhistoryintergalactic}, the obtained values of \(\tau\) were then rescaled to match different mean transmission values \(\bar{F}\) to compensate for our lack of knowledge of the \ac{UVB} amplitude.
Generally this rescaling will affect the shape and large scale amplitude of the power spectrum. \citet{lukic2015Lymanforestoptically} investigated this issue (see their Figure 23) and found that rescaling \(\tau\) by a factor of \(\sim 0.5\) results in to \(\sim 5\%\) changes in the \ac{Lya} forest power spectrum,
especially at low redshifts. While rescaling  \(\tau\) could be slightly biasing
our results, we emphasize that the rescalings we perform in this work are typically smaller \(\Delta\tau/\tau\sim 30\%\), and hence
this effect should be subdominant compared to e.g. boxsize effects and cosmic variance (see \autoref{sec:discussion}).

For each redshift and each parameter combination \(\vect{\Theta} =\{T_0,\gamma,\lambda_\mathrm{P},\bar{F}\}\) we generated \(50000\) randomly selected skewers -- the same ones for each parameter combination -- which serves as the starting point of our analysis.

\section{Measuring the Thermal State of the IGM} \label{sec:method}
In this section, we describe how we perform inference on our data using the THERMAL grid.
This involves generating a forward model of the data,
creating an emulator -- a fast method to interpolate from a sparse grid of simulation to any point in the multi-d parameter space, and finally performing the actual inference via Bayesian methods.

\subsection{Forward Modeling}

To compare to existing measurements, which didn't apply masking of spectral regions, but instead treated metal contamination statistically by comparing to lower redshift data where most metals are outside the \ac{Lya} forest, we compute the power spectrum based on \(\sim 50000\) noiseless, high-resolution skewers from our simulation. We will refer to this as the `perfect model'.

However, due to fully account for the window function introduced on the power spectrum by masking parts of the data, when comparing to our measurement from \citetalias{Walther2017NewPrecision}, we compute the power spectrum based on the skewers for each combination of parameters applying the full forward modeling technique described in \citetalias{Walther2017NewPrecision} to our hydrodynamical simulations.
Henceforth we'll call this the `forward model'.
This technique consists of several steps of post-processing the hydrodynamical simulation outputs followed by a power spectrum computation in the same way as for the data.
To forward model an individual quasar spectrum we first merge randomly selected skewers (without repetition) to cover the same pathlength as the data, then convolve the spectra with a Gaussian smoothing kernel reducing the resolution of the models to match that of the data, rebin the models onto the pixels of the observed spectra, and add noise drawn from a Gaussian distribution for each individual pixel with a standard deviation equal to the \(1\sigma\) uncertainty of the corresponding quasar spectral pixel reported by the data reduction pipelines.
Finally and most importantly, we mask the forward modeled spectrum in exactly the same way as the data to account for the windowing effects resulting from gaps in the data and our metal masking procedure.
We then compute the power spectrum by utilizing \(\sim 50000\) skewers from our simulation (see there \citetalias[see][for a more detailed description of the individual steps]{Walther2017NewPrecision}.
Note that while the full forward modeling of noise and resolution might not be completely necessary as they have been corrected in the measurement (and are corrected in the same way inside the forward modeling procedure as well), there might be subtle effects on the masking correction.
We therefore want to make the model spectra as similar to data as possible.
Note that this does not change our model precision which is dominated by dataset size rather than noise or resolution.

\subsection{Emulation of the Power Spectrum}\label{sec:emulator}
\begin{figure}
\plotone{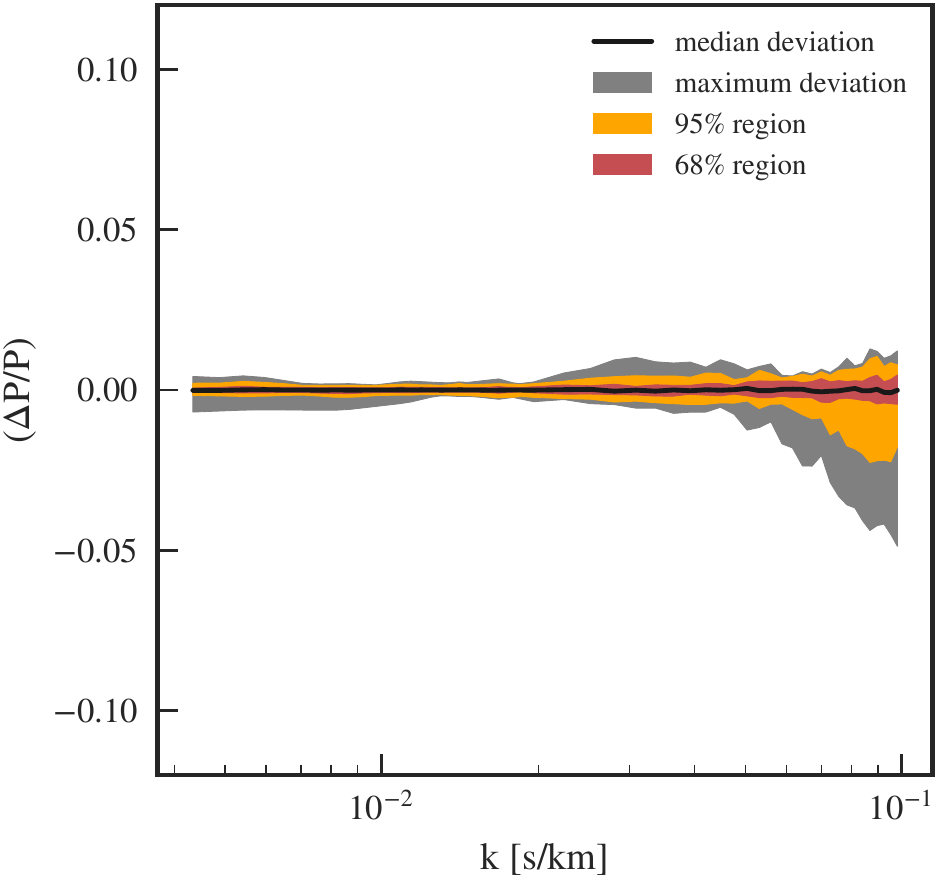}
\caption{Cross validation results for our emulation procedure at \(z=2.8\). Colored bands are showing the relative difference between emulated and true power for different cuts of the full cross validation set. The median is shown as a black curve. Other redshifts give similar results especially for the \(68\%\) region. See main text for more details.}
\label{fig:emu_test}
\end{figure}

To perform a fit to the data and infer the thermal state at a particular redshift we need to be able to compute power spectra on a continuous range of parameters.
Therefore we need to interpolate between the discrete and sparse outputs of the THERMAL grid.
To perform this task we follow the emulation approach of \citet{heitmann2006Cosmiccalibration} and \citet{habib2007CosmicCalibrationConstraints}. For details, we refer the reader to their papers (and references therein) as well as \citetalias{Walther2017NewPrecision}; in the following we summarize the main steps of the approach.
First, we decompose the simulated logarithmic power spectra onto a \ac{PCA} basis.
We save the PCA vectors as well as the coefficients $A_i(\vect{\Theta}_j)$ at each thermal model location $\vect{\Theta}_j$. We then use a Gaussian process to interpolate the coefficients $A_i(\vect{\Theta}_j)$ onto any arbitrary location in parameter space $\vect{\Theta}$. Taking the dot product of the PCA vectors with these interpolated coefficients then gives the power spectrum evaluated at any parameter location.

We thus calculate a \ac{GP} for each principal component coefficient \citep[using \texttt{GEORGE}, see][]{ambikasaran2016FastDirectMethods} using a squared exponential kernel plus an additional white noise contribution
\begin{equation}
K(\vect{\Theta}_1,\vect{\Theta}_2)= \exp(-0.5(\vect{\Theta}_1-\vect{\Theta}_2)C_l^{-1}(\vect{\Theta}_1-\vect{\Theta}_2)) + \sigma_n \delta_\mathrm{ij}
\end{equation}
for parameter values \(\vect{\Theta}_i\), a chosen distance metric \(C_l\) (which is defined by a smoothing length \(l\) for each parameter, i.e. it's diagonal) and a noise contribution \(\sigma_n\) (for an in depth introduction to \ac{GP} techniques, see \citealt{Rasmussen2005GaussianProcesses}).

As the hydrodynamical grid consists of far less models (\(\sim 50\))\footnote{The exact number of models used is redshift dependent because of further cuts that are discussed at the end of this subsection.} than the previous \ac{DM} based grid (\(\sim 500\)) used in \citetalias{Walther2017NewPrecision}, we must be more careful about the interpolation errors resulting from our emulation procedure.
Instead of just using a kernel with a fixed hand-tuned smoothing length, which was our approach in \citetalias{Walther2017NewPrecision}, we additionally optimized our kernel parameters by maximizing \ac{GP}-likelihood using the \texttt{scipy.optimize} \citep{Jones2001SciPy:Open} package and the so-called \texttt{L-BFGS-B} \citep{Zhu1997Algorithm778:} method\footnote{If a low likelihood was achieved we optimized again using the downhill simplex method by \citealt{Nelder1965SimplexMethod} and took the more optimal of the 2 runs}. We then performed the analysis using the optimal smoothing lengths \(l\) and noise \(\sigma_n\)
for the kernel for each Gaussian process emulator.

We estimate the emulation uncertainties using a cross-validation scheme to propagate interpolation errors. To do this we generate the emulator, but leave one simulation out of the training set\footnote{In fact we discard all the different \(\bar{F}\) realizations for this simulation in this test as they all have the same thermal parameters}. We denote emulators with a model (defined by parameters \(\vect{\Theta}\)) left out as \(\mathrm{emu}\backslash \vect{\Theta}\).
We then compare the actual models (with power \(P_\mathrm{model}\)) for this simulation to the emulator (with power \(P_{\mathrm{emu}\backslash \vect{\Theta}}\)) at the parameters \(\vect{\Theta}\) of this model:
\begin{equation}
  \Delta P_\mathrm{emu}(\mathbf{k},\vect{\Theta})=P_\mathrm{model,\vect{\Theta}}(\mathbf{k})-P_\mathrm{emu\backslash \vect{\Theta}} (\mathbf{k},\vect{\Theta}).
\end{equation}
We show the accuracy of the emulation in \autoref{fig:emu_test}.
This shows quantiles of the deviations \(\Delta P_\mathrm{emu}\) from the true underlying model inside our cross-validation sample.
We see that for most models in our parameter space the emulator works to better than \(1\%\). However, emulation uncertainty can increase to the \(5\%\) level (with a preference for underestimation at \(k>\SI{0.06}{\skm}\)) for some models.
As the uncertainty in our power spectrum measurements is \(\sim 2 \%\) (for the \(68\%\) quantile) on large scales (\(k\lesssim \SI{0.01}{\skm}\)) and \(\gtrsim 5\%\) on smaller scales, measurement errors are much larger than these interpolation errors.
Nevertheless, we opted to add the covariance matrix for the interpolation process to our likelihood.
This covariance matrix can be obtained by performing:
\begin{equation}
  C_\mathrm{emu,ij}=\langle \Delta_{\mathrm{emu}}(\mathbf{k_\mathrm{i}},\vect{\Theta})\Delta_\mathrm{emu}(\mathbf{k_\mathrm{j}},\vect{\Theta}) \rangle
\end{equation}
with the average performed over all possible combinations of model parameters inside our grid for each redshift bin.

Due to the variety of thermal histories in the THERMAL suite some simulations can have extremelly close values of their thermal parameters at some specific redshifts. In order to avoid possible problems in the emulator due to this issue we removed models from the THERMAL grid that did not satisfy a distance threshold\footnote{To be precise we demand \(\sqrt{\sum_\mathrm{\Theta \in \vect{\Theta}}\left(\frac{\Theta_i-\Theta_j}{\max(\Theta)-\min(\Theta)}\right)^2}\geq 0.1\). As we have about 10 bins in \(\gamma\), and about 7 in \(T_0\), the separation between adjacent points would be at least 0.1 in the units shown. But no two models have the same \(\lambda_P\), increasing the separation. Therefore, the chosen minimal separation is still closer than our typical grid separation. While this threshold leads to good results throughout our redshift range, it is not necessarily the optimal one and further tests adopting different values could therefore be used to slightly increase interpolation accuracy.} and are left with \(45\) to \(65\) models per redshift.

\subsection{Inference}
\label{sec:inference}
We perform a Bayesian \ac{MCMC} analysis on the power spectrum data at each individual redshift using the \texttt{emcee} package \citep{foreman-mackey2013emceeMCMChammer} based on the affine invariant sampling technique \citep{goodman2010Ensemblesamplersaffine} and assuming the multivariate Gaussian likelihood:
\begin{align}
\mathcal{L}\equiv&P(\mathrm{data}|\mathrm{model})\\
\propto&\prod_\mathrm{datasets}\frac{1}{\sqrt{\det(C)}}\exp\left(-\frac{\mathbf{\Delta}^\mathrm{T} C^{-1} \mathbf{\Delta}}{2}\right) \nonumber\\
\mathbf{\Delta}=&\mathbf{P}_\mathrm{data}-\mathbf{P}_\mathrm{emu} \nonumber\\
C=&C_\mathrm{data}+C_\mathrm{emu}. \nonumber
\label{eq:likelihood}
\end{align}
with \(C_\mathrm{emu}\) being the covariance of the interpolation procedure and \(C_\mathrm{data}\) being the covariance of an individual measurement.
For these covariances we use published values if available. For our own dataset from \citetalias{Walther2017NewPrecision} as well as the \citet{viel2013Warmdarkmatter} dataset, we used the published uncertainties (i.e. the diagonal covariance elements) and combined them with the correlation matrix of the model closest in parameter space to obtain an estimate of the covariance, i.e. we perform nearest neighbor interpolation between covariance matrices obtained at every point (see \citetalias{Walther2017NewPrecision} for details on this approach).

\subsection{Parameters and Priors} \label{sec:priors}
\begin{figure}
\centering
\plotone{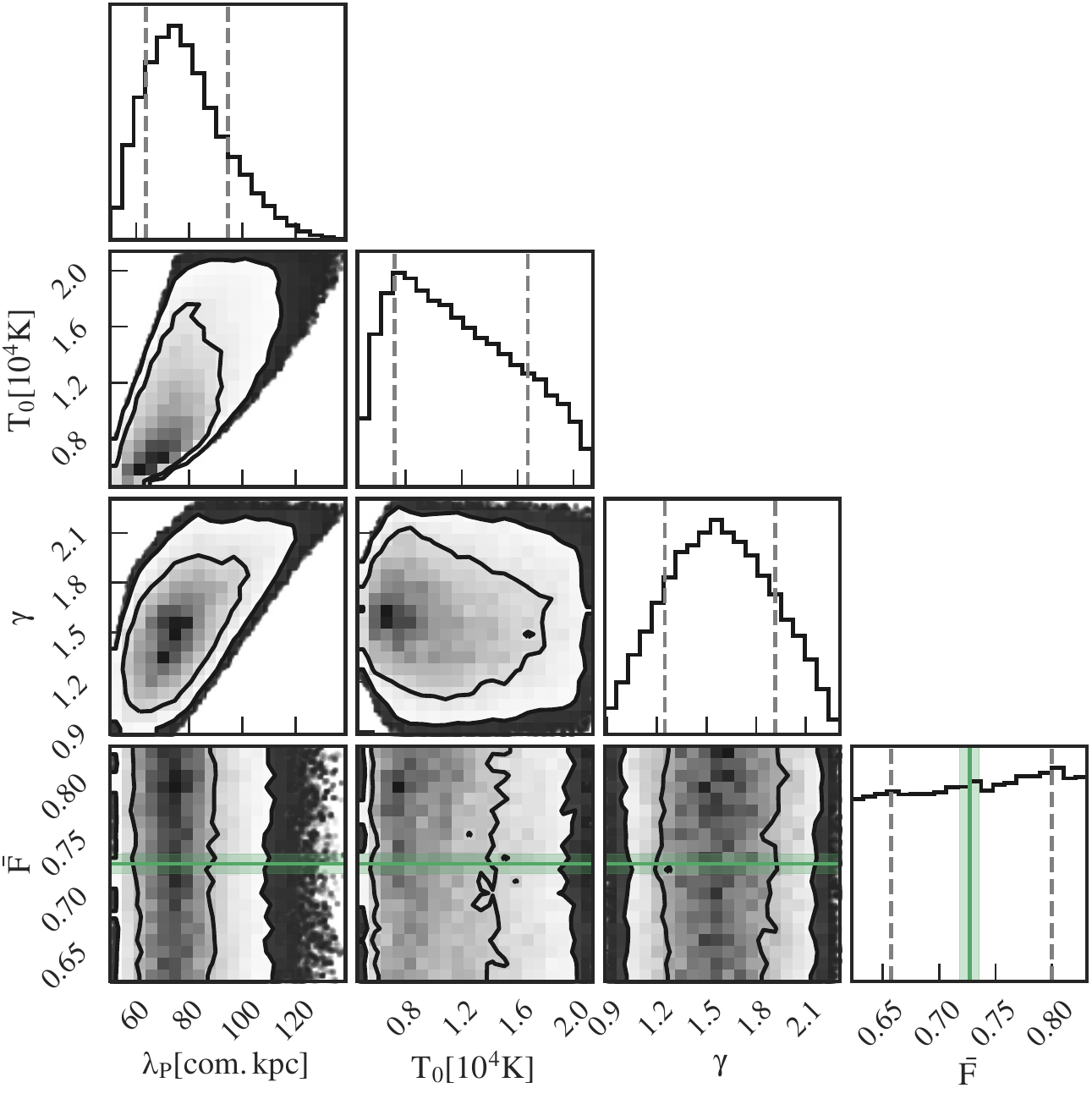}
\caption{Corner plot showing the Prior PDF for thermal parameters and \(\bar{F}\) given our model grid and excluding parameter values outside its convex hull. This was obtained by sampling our prior with an MCMC assuming a flat likelihood. Note that the degeneracies in our model grid lead to non-flat marginalized distributions.
The diagonal shows the 1d-PDF (marginalized over all other parameters) for each parameter with dashed vertical lines at the \(16\%\) and \(84\%\) quantiles. The scatter plots below show the 2d-PDFs for each combination of 2 parameters (also marginalized over all others) with contours showing the region containing the \(68\%\) and \(95\%\) highest densities.
Note that due to the restrictions of our grid there is a strong correlation especially between \(T_0\) and \(\lambda_\mathrm{P}\). The additional preference towards low \(T_0\) or \(\lambda_\mathrm{P}\) is due to our choice of flat priors in the \(\log\) of these parameters. The green band shows the \(1\sigma\) interval in \(\bar{F}\) we use for the Gaussian prior.}
\label{fig:prior}
\end{figure}

Our modeling so far depends on 4 parameters, \(T_0\) and \(\gamma\) describing the thermal state, \(\lambda_\mathrm{P}\) for the pressure smoothing depending on the full thermal history, and \(\bar{F}\) for the mean transmission that corresponds to a given UVB amplitude.
There is, however, one additional parameter that we input in our models for each dataset\footnote{As the treatment of metal lines in the different data sets is following fundamentally different approaches, masking which should remove at least part of the \SiIII{} in the spectra vs. subtraction of the metal power estimated from side bands, we decided to allow a different value for each measurement. Note that in the end thermal parameters will not be strongly correlated with the \SiIII{} parameter.} to generate the observed correlation between
\SiIII{} and \ac{Lya} \citep[see][]{mcdonald2006LyForestPower,palanque-delabrouille2013onedimensionalLy}.
Finally, because of significant uncertainties in the resolution of the XQ-100 data
(see the detailed discussion in Appendix B of \citetalias{Walther2017NewPrecision}),  we also marginalize over the resolution of the XQ-100 measurement whenever we use this data, giving us another parameter.
Therefore we have a total of 5 (in the case of high-resolution data only) to 8 (in the case of fitting 3 datasets of which one comes from XQ-100) parameters.
We assume flat priors on \(\log T_0, \log \lambda_\mathrm{P}, \gamma\).
We now go into further detail about the modeling and assumptions for the other parameters.

We add \SiIII{} correlations to the model analytically by multiplying the model power spectrum with an oscillating signal as correlations inside a spectrum correspond to oscillations of the corresponding power spectrum:
\begin{equation}
P_\mathrm{tot}=(1+a_\mathSiIII^2+2a\cos(k\, \, \Delta v)) P_\mathHI
\label{eq:SiIII}
\end{equation}
with \(a_\mathSiIII\) being a free nuisance parameter for the strength of the correlation.
In previous works this was typically expressed as \(a_\mathSiIII=f_\mathSiIII/(1-\bar{F})\) with \(f_\mathSiIII\) being a redshift independent quantity that was fit using the entire
dataset. We adopt this same parametrization but opt to fit for a unique value of \(f_\mathSiIII\) at each redshift and for each dataset because of the different metal treatment in the datasets and as we do not perform a joint fit of different redshifts here.
We assume a flat prior on each \(f_\mathSiIII\) and demand correlations to be positive.

We modeled the resolution of the X-SHOOTER spectrograph \(R_\mathrm{new}\) by multiplying the measured XQ-100 power spectrum with the resolution dependent part of the window function:
\begin{equation}
W_R(k,R) = \exp\left(- \frac{1}{2}(kR)^2\right)
\label{eq:window}
\end{equation}
using the resolutions quoted in \citet{Irsic2017Lymana} and dividing by \(W_R(k,R_\mathrm{new})\).
Note that the resolution of the instrument depends on two different factors: the resolution for a fully illuminated slit (or "slit resolution") and the seeing which gives rise to higher spectral resolution if smaller than the slit size.
We assume two limits for the resolving power of the XQ-100 dataset. The lower limit assumes the slit resolutions quoted in the XQ-100 data release paper \citep{lopez2016XQ100legacysurvey} as well as a fully illuminated slit (leading to \(R_\mathrm{UVB}=4350\) for the UVB arm of the instrument, \(R_\mathrm{VIS}=7410\) for the VIS arm\footnote{These values are also close to the formerly quoted "new values" from the instrument website as well as manuals until Period 101.}). The upper limit assumes a seeing of \(\SI{0.65}{\arcsecond}\) (smaller than the slit) and higher values for the slit resolution\footnote{Based on our on estimates of XSHOOTER's resolution in \citetalias{Walther2017NewPrecision} which is also close to the recently updated values on the XSHOOTER website and manual from Period 102} (leading to \(R_\mathrm{UVB}=8230\) and \(R_\mathrm{VIS}=12184\)).
We assume a flat prior between these two limits. As \(z=3.6\) is using both spectral arms we use the lowest and highest of the 4 resolution values above as the limits here.
Note that this choice of priors on spectroscopic resolution is an extremely conservative choice that will significantly weaken the constraints that can be obtained from this XQ-100 dataset.
This is most acute in the UVB arm because of its intrinsically lower resolution.
A more careful analysis of the XQ-100 resolution would allow us to adopt a far stronger prior on these values, which would increase the
precision of constraints deduced from power spectra measured from such moderate resolution spectra.

Note that most previous measurements \citep[exceptions to this are e.g.][]{lidz2010MeasurementSmallscale,Irsic2017Newconstraints} of the IGMs thermal state did not attempt to marginalize over the uncertainty in the mean flux estimate. Instead, typically simulations that match the mean flux of the data assuming perfect knowledge of this quantity are used (e.g. in Voigt profile fitting or curvature analyses).
For \(\bar{F}\) we used both a flat prior (corresponding to performing a joint measurement of \(\bar{F}\) and the thermal state) and a Gaussian shaped prior.
For the Gaussian prior we assumed a mean based on the fit by \citet{Onorbe2017Self-consistentModeling} to a compilation of recent measurements \citep{fan2006Constrainingevolutionionizing,kirkman2007ContinuousstatisticsLy,faucher-giguere2008DirectPrecisionMeasurement,becker2013refinedmeasurementmean}
and a standard deviation based on the uncertainties for the most recent measurements at \(z\le 4.0\): \citet{becker2013refinedmeasurementmean} for \(2.2\leq z\leq 4.0\), \citet{faucher-giguere2008EvolutionIntergalacticOpacity} for \(z=2.0\), \citealt{kirkman2005HIopacityintergalactic} for \(z=1.8\). For \(z \geq 4.2\) we use \(\sigma_{\bar{F}}=0.03\) which is loosely based on the discrepancy between \citealt{fan2006Constrainingevolutionionizing} for \(z\geq4.6\) and the measurements by \citealt{becker2011DetectionextendedHe} in the range \(4.1\leq z\leq4.7\) \citep[see also][for more recent mean flux measurements that are discrepant by a similar amount  for \(5.0\le z\le 5.4\)]{Bosman2018Newconstraints,Eilers2018OpacityIntergalactic}.

To avoid extrapolating from our model grid we additionally require that all thermal parameters lie inside the convex hull of our model grid (see \autoref{fig:grid_singlez}), i.e. the smallest convex shape including all THERMAL grid points.
The convex hull is evaluated numerically by triangulating the model grid (using \texttt{scipy.spatial.Delaunay}) and for each MCMC sample we test whether it is inside the
triangulation when evaluating the prior. Otherwise the prior is set to zero.
To see the effective prior resulting from only using this non-rectangular region
where we have models,  we performed an MCMC run assuming a completely uninformative dataset, i.e. using only the priors in our fit and a constant likelihood.
The results of this procedure are shown in \autoref{fig:prior} for \(z=2.8\).
In some contours, e.g. \(T_0\) and \(\lambda_\mathrm{P}\), we can see that parameters are highly correlated already since our grid is non-rectangular.
We argue, however, that these correlations are physically motivated as models perpendicular to these correlations are hard to produce (see \autoref{sec:hydro}) and that this behavior actually constitutes prior information for our analysis.

\section{Thermal Evolution of the IGM}
\label{sec:results}

\subsection{Measurements and Degeneracies}
\label{sec:measurements-degeneracies}
\begin{figure*}
\centering
\plotone{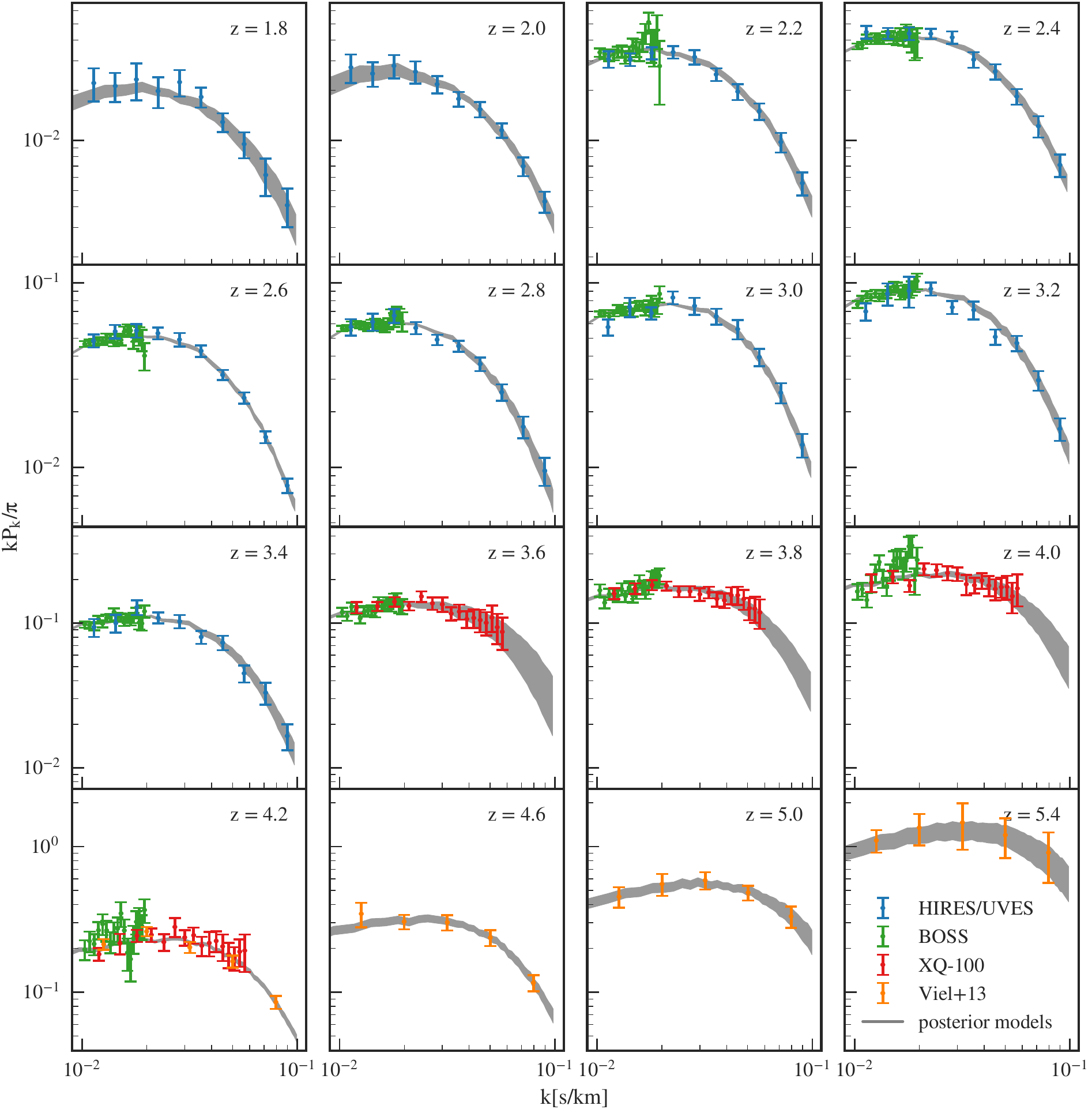}
\caption{Redshift evolution of the power spectrum with colors showing different datasets. Data by \citet{Irsic2017Lymana} were corrected to the median of the marginalized posterior resolution, \citet{Walther2017NewPrecision} points have been corrected for the masking window function. All data have been corrected for \SiIII{} correlations. Bands show \(68\%\) confidence regions for our emulator with parameters randomly drawn from the posterior distribution.
}
\label{fig:power_fit_singlez}
\end{figure*}
\begin{figure*}
\centering
\plotone{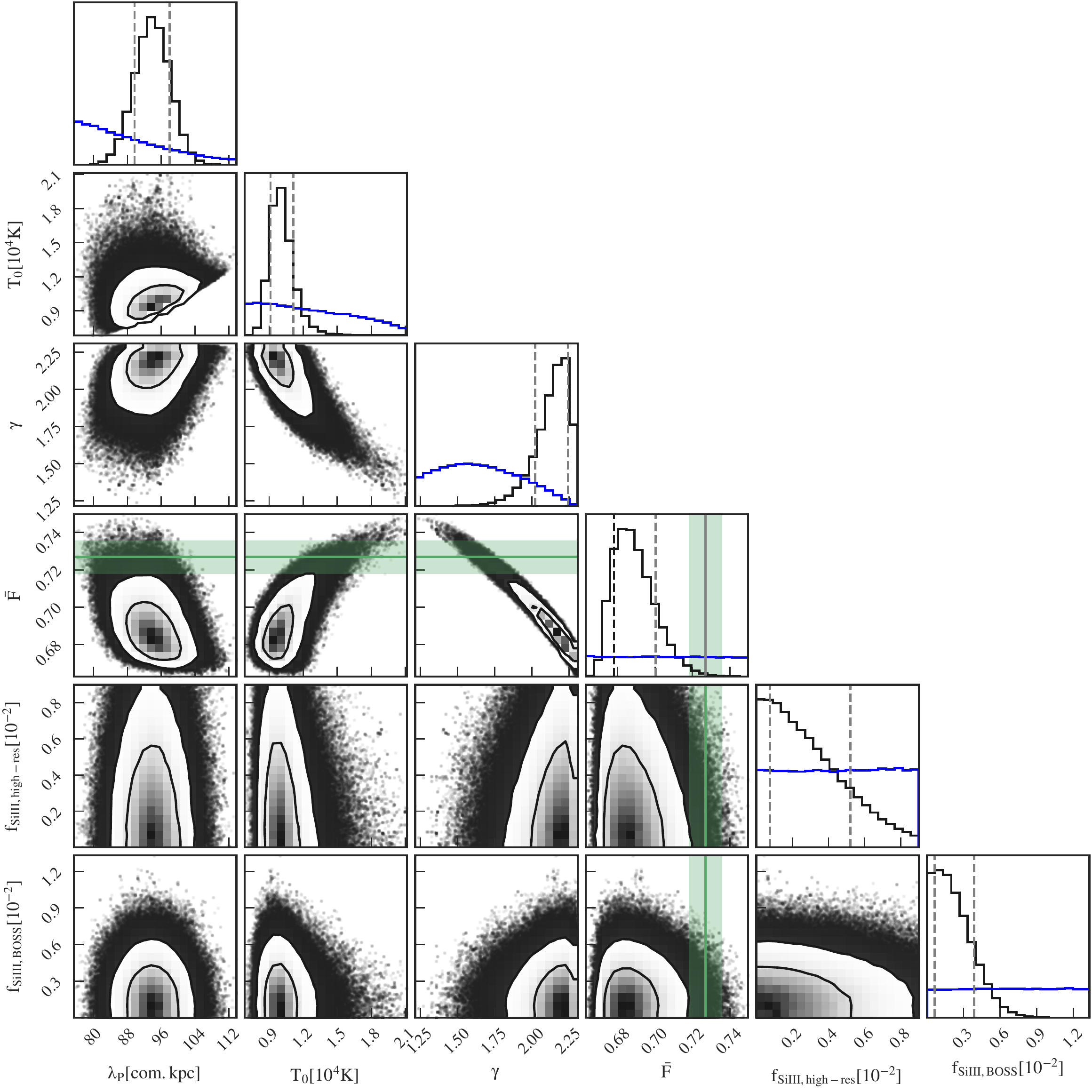}
\caption{Corner plot showing 1d- and 2d- marginalized posterior distributions for all fitting parameters at \(z=2.8\) assuming a flat prior on \(\bar{F}\). Blue curves in the 1d-histograms show 1d- marginalized distribution when ignoring the data and fitting the prior only (i.e. the result of the analysis performed for \autoref{fig:prior}).
We can see that there are strong constraints on all parameters compared to the prior information.
We also notice a strong correlation between permutations of \(\gamma\), \(T_0\) and \(\bar{F}\).
Note that the posterior in \(\bar{F}\) is significantly below the observed value of the \citet{becker2013refinedmeasurementmean} mean flux measurement (shown as a green line with a band for the 1-\(\sigma\)-region) which is, combined with the strong anticorrelation between \(\gamma\) and \(\bar{F}\) leading to higher values of \(\gamma\) than typically assumed.
}
\label{fig:corner_28}
\end{figure*}
We performed fits of the parameters governing the thermal state using combinations of all datasets discussed in \autoref{sec:measurements} in 16 individual redshift bins with \(1.8<z<5.4\), where we used a bin size \(\Delta z=0.2\) for \(z\le 4.2\) and \(\Delta z=0.4\) for \(z\ge 4.6\).

The power spectra of each dataset are summarized and compared to models based on our posterior MCMC chains in \autoref{fig:power_fit_singlez}.
Note that for visualization purposes we only compare window-function, \SiIII{} correlation and resolution corrected data
to the perfect model.
The window function due to masking was taken out of the UVES/HIRES data by multiplying measurement points with the median \(P_\mathrm{emu,perfect}/P_\mathrm{emu,forward}\) for our MCMC chain and propagating its uncertainties using Gaussian error propagation for each individual mode \citepalias[see][for a more detailed description of this process]{Walther2017NewPrecision}\footnote{Note that while we used DM models to correct the "raw" power in \citet{Walther2017NewPrecision}, the masking correction performed here is fully based on hydrodynamical simulations}.
Analogously, we rescaled the XQ-100 power to use the "best-fit" resolution correction, i.e. we renormalize with \(W_R(k,R_\mathrm{new})/W_R(k,R)\) (see \autoref{eq:window}) from the posterior and removed \SiIII{} correlations from the data applying \autoref{eq:SiIII}.
We can see that satisfactory fits have been achieved at all redshifts.

In \autoref{fig:corner_28} we further illustrate the posterior
distribution we infer via our MCMC at $z=2.8$ with a so-called `corner plot'.  We
can see that the data strongly constrains all parameters (e.g. compare
to \autoref{fig:prior} or the blue curves in the 1d histograms, for which the likelihood is assumed to be completely uninformative). The most important feature we see is that there are
strong degeneracies between some parameters, e.g. the diagonal
contours between permutations of \(T_0\), \(\gamma\) and \(\bar{F}\).
Note that the strong correlation between \(T_0\) and \(\gamma\) is well understood and results from the IGM not probing the mean density, but instead mild overdensities at these redshifts \citep[see e.g.][]{lidz2010MeasurementSmallscale, becker2011DetectionextendedHe}.
We also infer a low mean transmitted flux \(\bar{F}= 0.69\pm0.01\)
compared to the \citet{becker2013refinedmeasurementmean} measurement
of \(\bar{F}= 0.727\pm 0.009\) (green band). It is interesting to note
that this low value however agrees well with the joint constraint on mean transmission evolution
by \citet{Palanque-Delabrouille2015Neutrinomasses} obtained from the
BOSS power spectrum yielding \(A=0.0028\pm0.0002,\eta=3.67\pm0.02\) for \(\bar{F}(z)=\exp(-A (1+z)^\eta)\) resulting in  \(\bar{F}(z=2.8)\approx 0.687\pm0.020\).
Note that the dataset used in this analysis overlaps with the one we used here, but simulations and inference procedure are independent and our analysis has additional higher resolution data available.
Independent of the BOSS data, we also obtain similarly low \(\bar{F}\)
values when performing fits on the high-resolution data from
\citetalias{Walther2017NewPrecision} alone.

Additionally, the posterior distribution for \(\gamma\) shows a clear preference
for  values \(\gamma\approx 2.1\), far above the expected value of \(\sim 1.6\) for
IGM gas in photoionization equilibrium long after reionization events
\citep{hui1997Equationstatephotoionized,McQuinn2016intergalactictemperature-density}
Note again that there is a strong anti-correlation
between \(\gamma\) and \(\bar{F}\), so while our analysis prefers
a  high value of \(\gamma\) and a low value of \(\bar{F}\), this is a movement along the degeneracy direction.  We will further discuss this issue in \autoref{sec:discrepancies}.

\begin{figure*}
\centering
\plotone{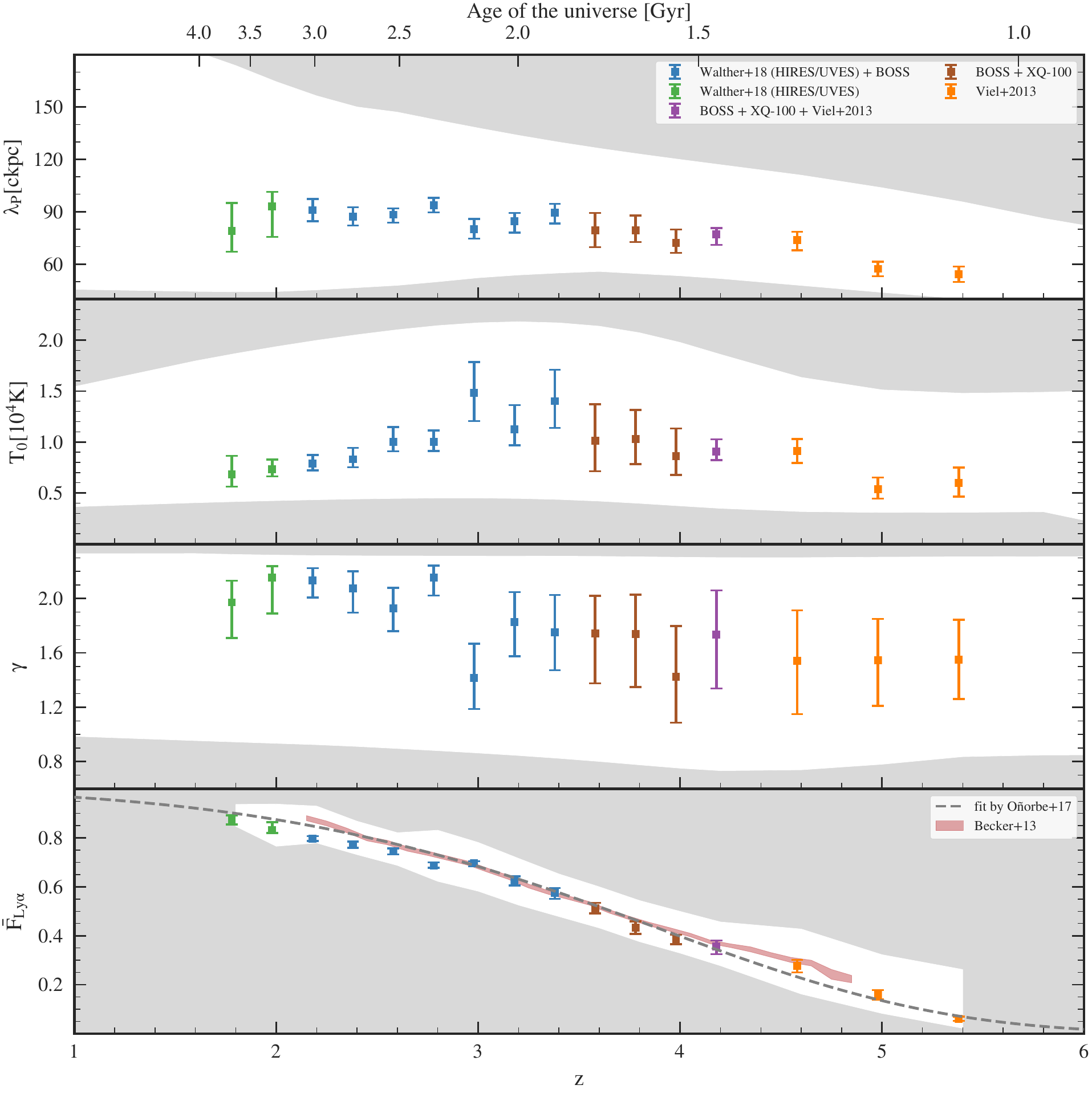}
\caption{ Points with error bars show the median and the region between \(16\%\) and \(84\%\) quantiles of the 3 thermal parameters as well as the mean transmission of the IGM (marginalized over all model parameters of the fit) at different redshifts using our fiducial dataset (squares) at each redshift.
In the \(\bar{F}\) panel we also show the evolution obtained by
\citet{becker2013refinedmeasurementmean} (red band showing \(1\sigma\)
uncertainties) based on relative changes in SDSS quasar transmissions
as well as the \citet{Onorbe2017Self-consistentModeling} (dashed line)
fit to these data and further datasets.  Note the large discrepancies
between our measurements and those results when assuming an
uninformative prior on the mean flux.  The white range shows the space
populated by our models, i.e. we cannot expect to measure values
inside the gray shade using our current emulator. }
\label{fig:evolution_onlydata}
\end{figure*}

The redshift evolution of individual parameters, determined from the 1d marginalized posteriors, is illustrated in \autoref{fig:evolution_onlydata}. For \(3.0\leq z \leq 3.4\) we also performed fits including the XQ-100 data, and fully marginalized over our lack of knowledge of the exact spectroscopic resolution (see discussion in \autoref{sec:priors}).
As including this dataset did not significantly change our results, we decided to leave those points off the plot for clarity.
Numerical values for the marginalized parameters are tabulated in \autoref{tab:results_flat} in the Appendix.

There are several noteworthy features in \autoref{fig:evolution_onlydata}.
First, the disagreement that we saw at $z=2.8$ between our inferred value of \(\bar{F}\) and recent measurements is also present at all other redshifts \(z<3\) (green and blue datapoints compared to the pink shaded region in the lower panel).
At the same time \(\gamma\) reaches very high values in the same redshift range. Also \(T_0\) drops strongly from \(z=3.0\) to lower redshifts, but due to the degeneracies between \(T_0\), \(\gamma\), and \(\bar{F}\) these measurements are all strongly correlated and this effect is therefore expected.
Note that these trends -- high \(\gamma\), low \(\bar{F}\), and low \(T_0\) --
persists if we fit the high-resolution data alone, as the BOSS data alone do not individually constrain all of these parameters due to the lack of high-\(k\) modes (resulting from limited spectral resolution).

Second, for \(z\geq 3\) we can see that \(\gamma\) shows little evolution and the mean transmitted flux \(\bar{F}\) is consistent with the \citet{Onorbe2017Self-consistentModeling} fit to recent measurements.
We can also see that \(T_0\) increases from
\(\approx \SI{5100}{\kelvin}\) at \(z=5.0\) to \(\approx \SI{15000}{\kelvin}\) at \(z = 3.4\).
This rise could be explained by the onset of \HeII{} reionization, which we discuss in more detail in \autoref{sec:model_comp} where we compare our inferred parameter values to models of IGM thermal history that treat reionization heating.

In summary, we can see that the power spectrum analyzed here can in principle achieve high precision  constraints on IGM thermal parameters and the mean transmission, but the high values of \(\gamma\simeq 2\) inferred at \(z<3\) and concomitant discrepancies between our inferred mean flux and the \citet{becker2013refinedmeasurementmean} measurements
might indicate systematics in our procedure. We consider this issue in detail in the next section.

\subsection{Analyzing the Discrepancies in \(\gamma\) and \(\bar{F}\)}\label{sec:discrepancies}

\begin{figure}
\centering
\includegraphics[width=\columnwidth]{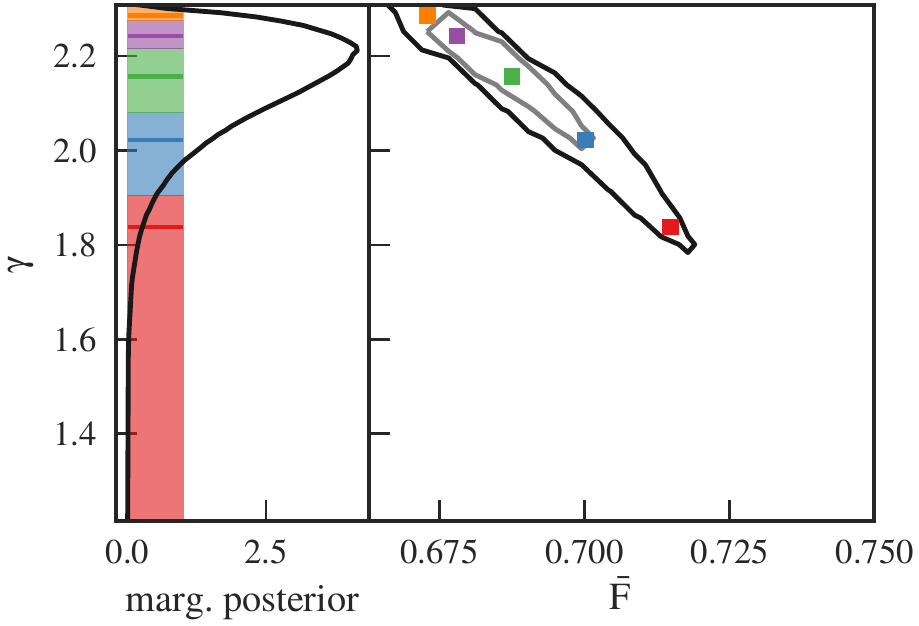}
\caption{Illustration for our approach in selecting models along the posterior distribution (see main text for details). Left:  The marginalized posterior distribution of \(\gamma\) values from our chain with the bins which we used to select generate models at the \(68\%\) and \(95\%\) confidence intervals shown as bars. The median chain value in each bin is shown as a colored line. Right: \(68\%\) and \(95\%\) contours for \(\gamma\) vs. \(\bar{F}\) with the selected values of both parameters shown as squares.
}
\label{fig:gammacomp_illustration}
\end{figure}

\begin{figure}
\centering
\plotone{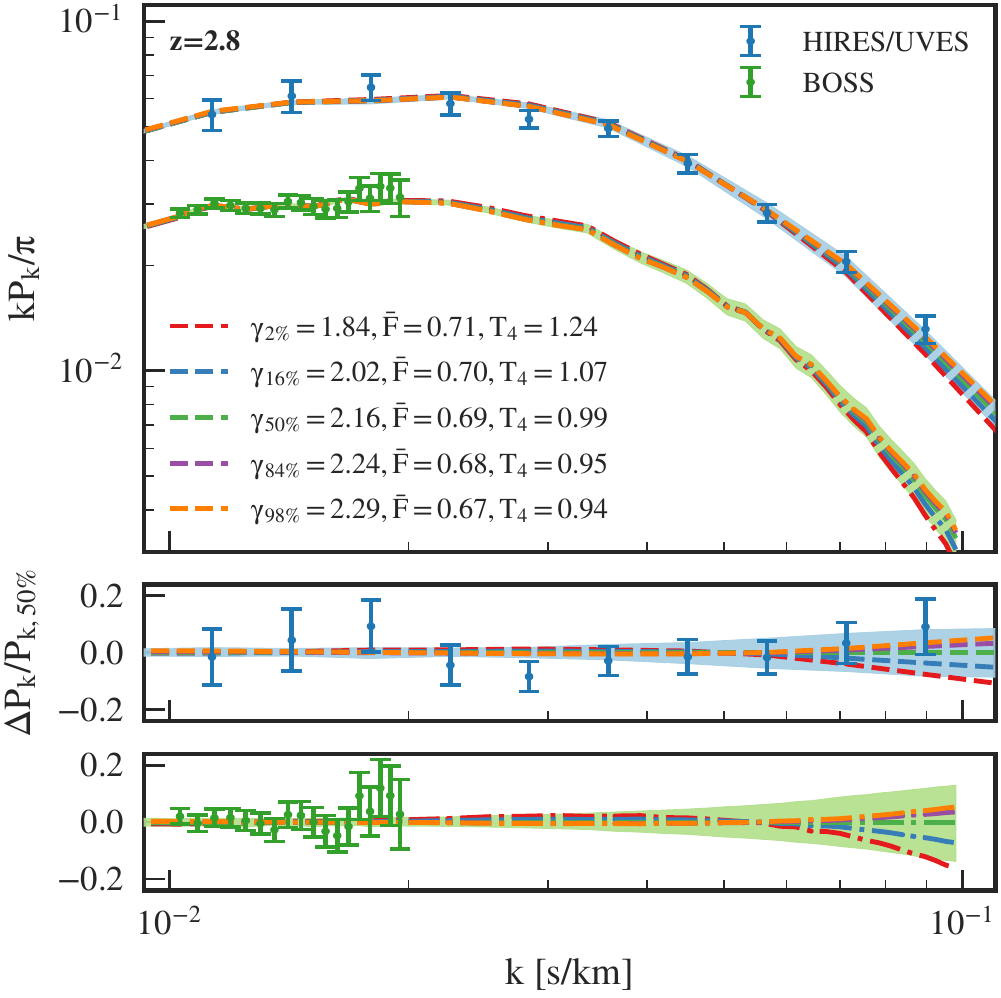}
\caption{Topmost panel: The power spectrum (not corrected for masking) at \(z=2.8\)  (other redshifts are shown in \autoref{fig:gammacomp_all}, bands are showing regions in which \(68\%\) of models in the posterior fall) with curves showing models (drawn from the respective emulator) with different thermal parameters. Those are chosen such that the lines represent the \(2.5\%, 16\%, 50\%, 84\%\) and \(97.5\%\) quantiles of the posterior distribution in \(\gamma\) while following degeneracies with the other parameters (see main text and \autoref{fig:gammacomp_illustration} for the details). Values of the most relevant parameters are printed inside the figure (with \(T_4=T_0/\SI{1e4}{K}\)). Both datasets have been offset by a factor of two for clarity.
Bottom panels: The fractional deviation between data in the topmost panel and the model at median \(\gamma\) (green curve) for each dataset.
}
\label{fig:gammacomp_28}
\end{figure}
\begin{figure*}
\centering
\includegraphics[width=0.3\textwidth]{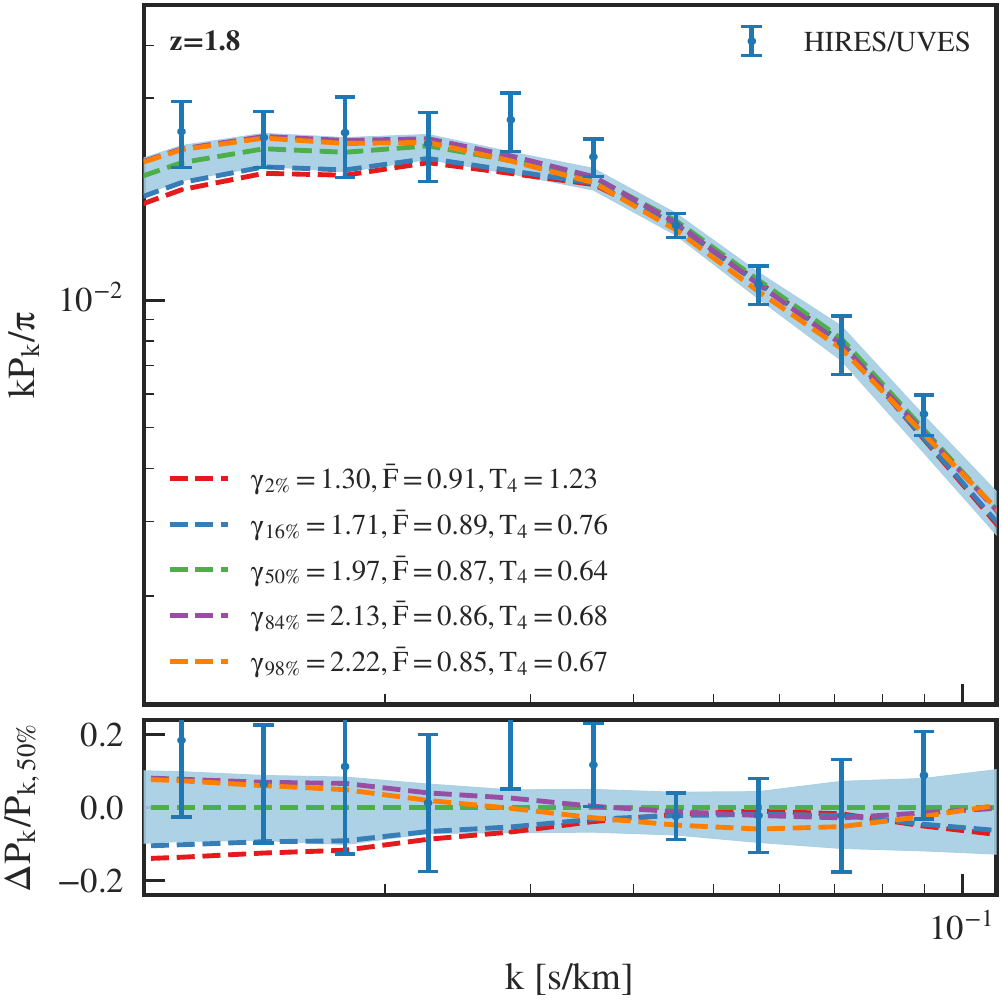}\hspace{0.3cm}\includegraphics[width=0.3\textwidth]{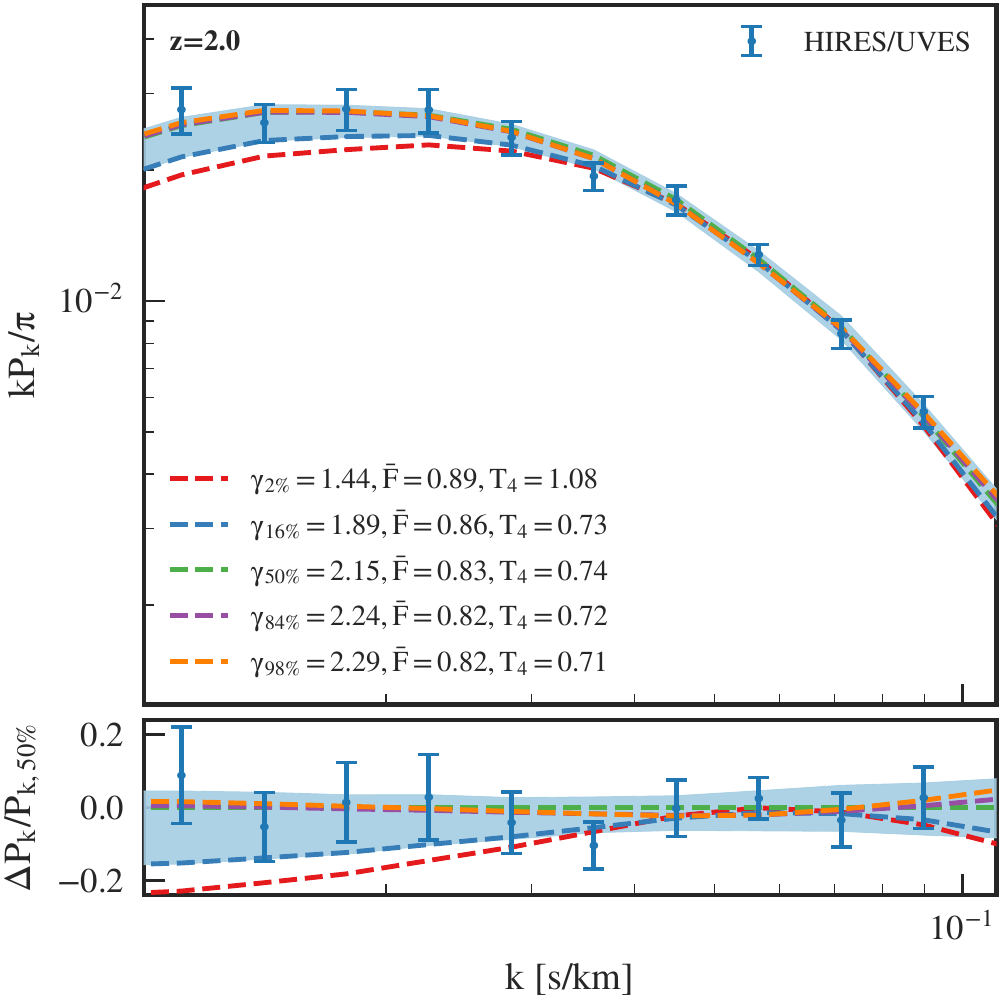}\hspace{0.3cm}\includegraphics[width=0.3\textwidth]{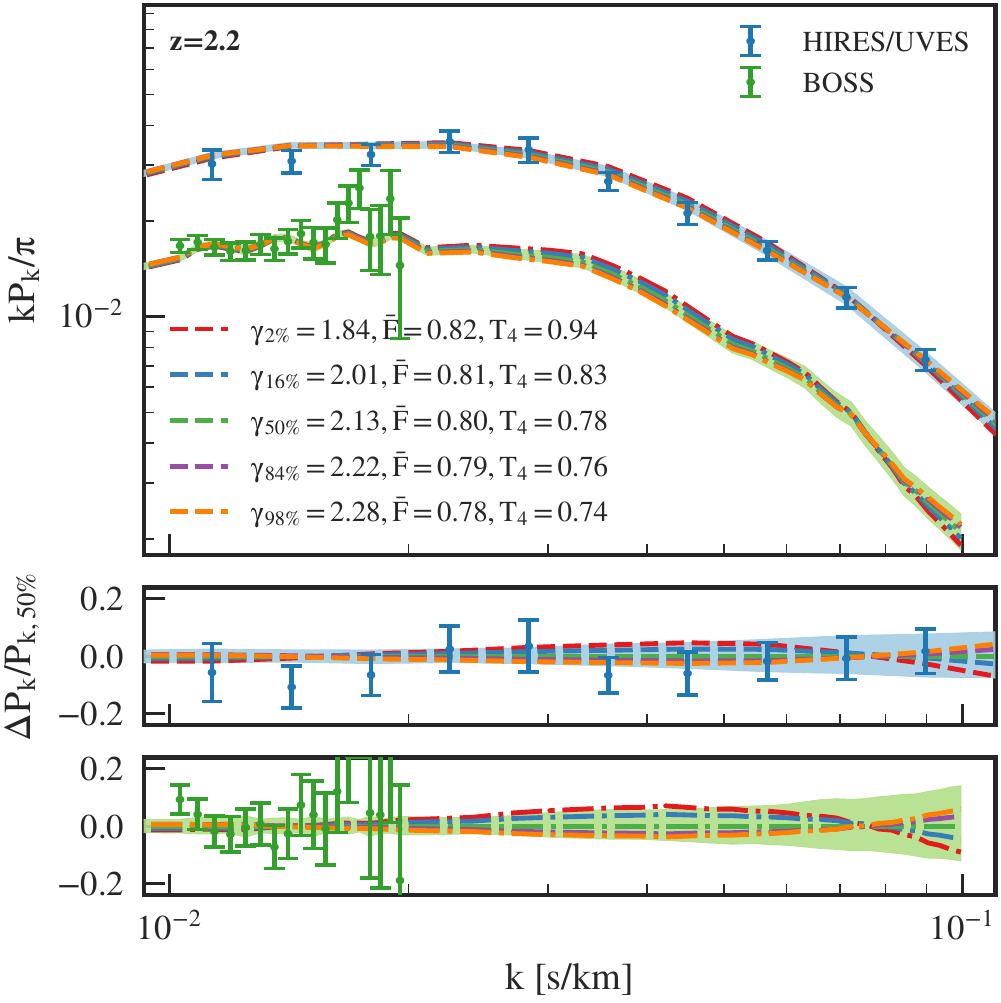}\\\vspace{0.3cm}
\includegraphics[width=0.3\textwidth]{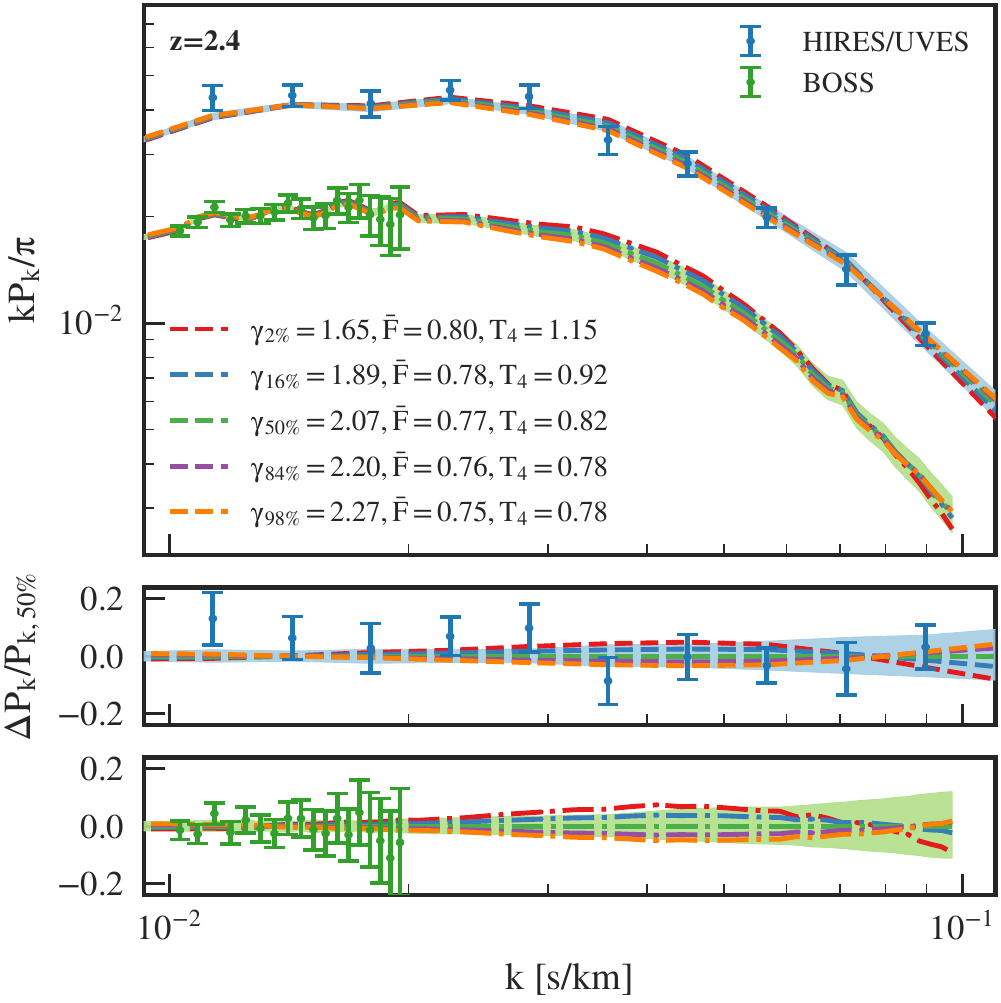}\hspace{0.3cm}\includegraphics[width=0.3\textwidth]{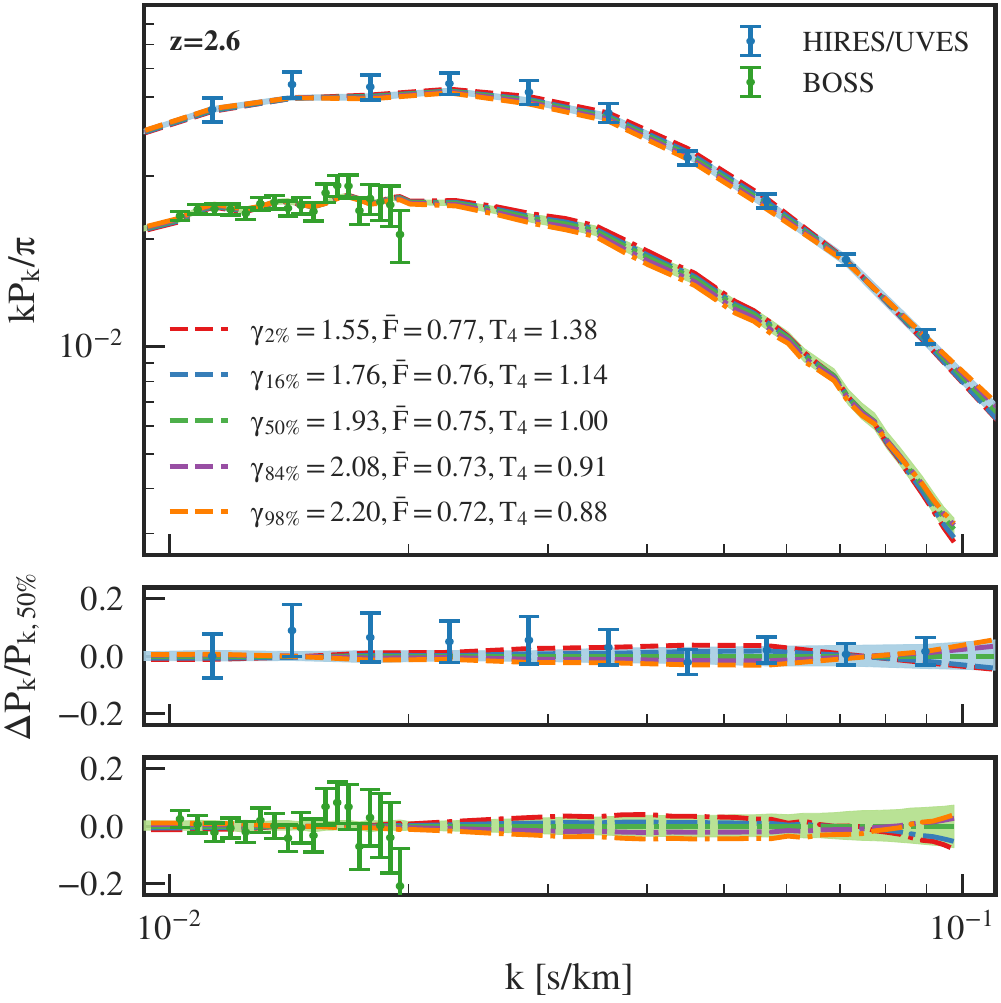}\hspace{0.3cm}\includegraphics[width=0.3\textwidth]{gamma_comp_z2_8}\\\vspace{0.3cm}
\includegraphics[width=0.3\textwidth]{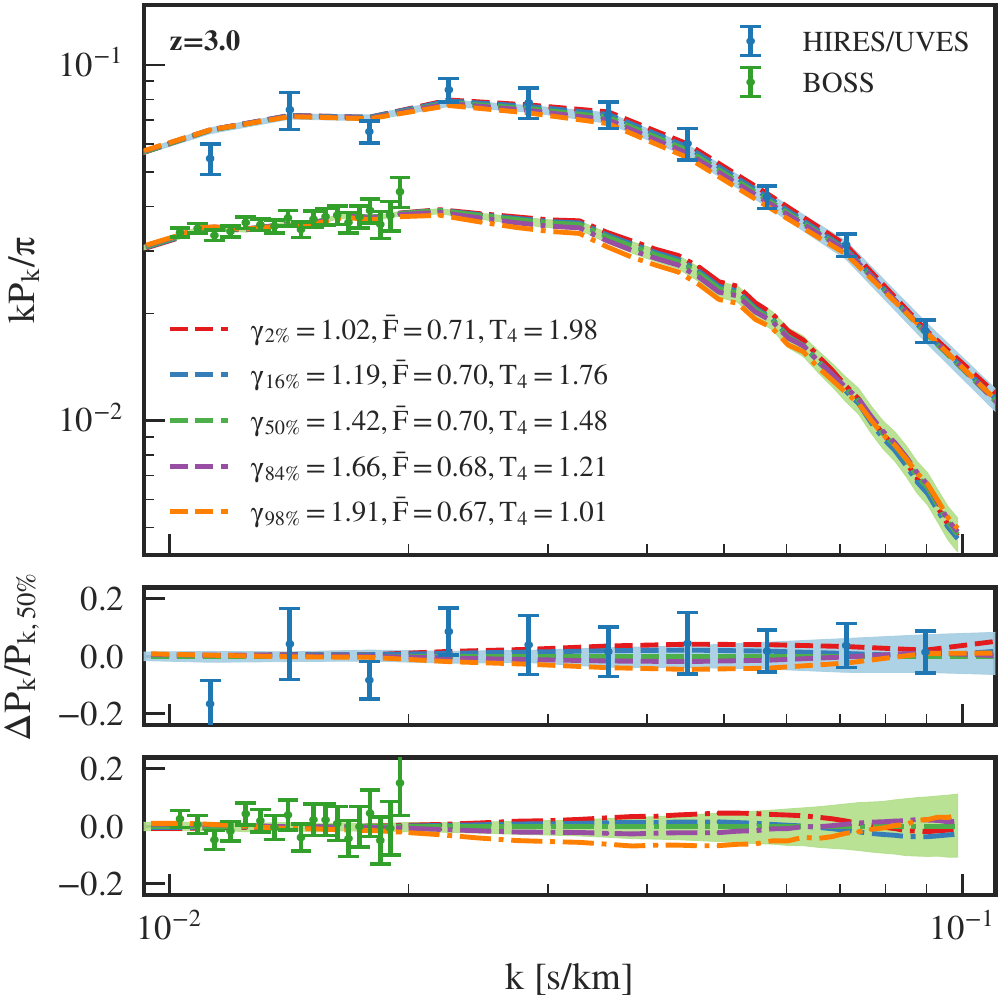}\hspace{0.3cm}\includegraphics[width=0.3\textwidth]{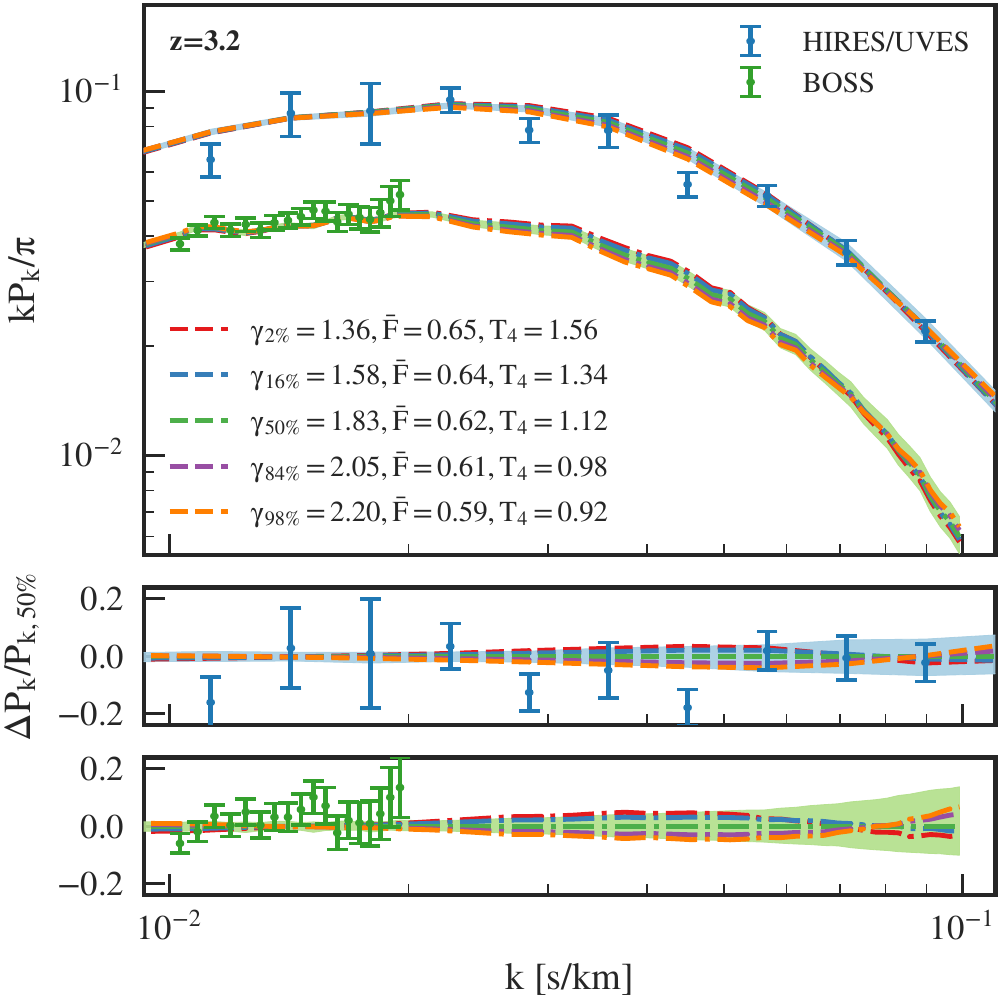}\hspace{0.3cm}\includegraphics[width=0.3\textwidth]{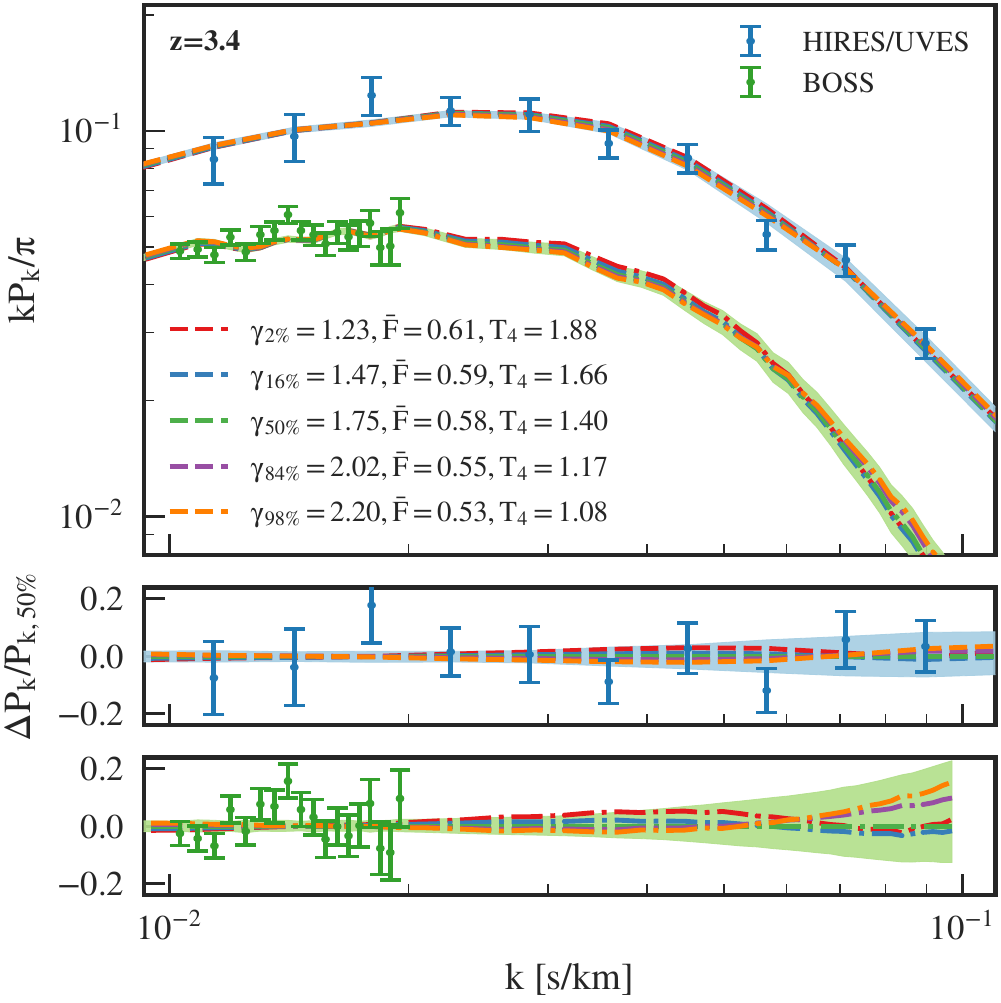}
\caption{Same as \autoref{fig:gammacomp_28}, but also for all redshifts \(z\leq3.4\). We can see that while most redshift bins show the strongest scatter in the power at \(k\sim \SI{0.06}{\skm}\) when moving along the degeneracy direction. However, for \(z=1.8, 2.0\) the behaviour seems to be significantly different most likely due to the lacking precision on small \(k\) due to the lack of the BOSS measurement at these redshifts.
}
\label{fig:gammacomp_all}
\end{figure*}

In the previous section we found low values of \(\bar{F}\) compared to \citet{becker2013refinedmeasurementmean} and possibly unphysically high values of \(\gamma\).
While both parameters are degenerate and the degeneracy direction matches with our discrepancy this might point towards some problem within the analysis.
To investigate this scenario we want to isolate the change in the power spectrum when moving along the degeneracy direction of our posterior distributions.
Due to the dimensionality of the parameter space and correlations between different parameters this can't be achieved by simple cuts along a parameter direction.
Therefore we designed the following procedure to generate model curves tracking the degeneracy direction for different values of \(\gamma\) (also see the illustration in \autoref{fig:gammacomp_illustration}):
\begin{itemize}
\item We take the posterior of our MCMC analysis (i.e. the Markov chain) and define bins such that the median of \(\gamma\) inside a bin is equal to a desired quantile of the marginalized \(\gamma\) distribution (which are chosen to be equivalent to \(\pm1\sigma\),\(\pm2\sigma\)).

  These bins are shown as colored bars in the left panel of \autoref{fig:gammacomp_illustration}.
\item For $\gamma$ values in our chain within a given bin,  we then compute the median of all other parameters. Because of the way we chose our $\gamma$ bins, this yields the quantile of interest for \(\gamma\), whereas
  the other parameters will track their corresponding degeneracy direction with respect to \(\gamma\). This can be seen in the colored squares in the right panel of \autoref{fig:gammacomp_illustration}.
\item For the set of parameters at each of the quantiles (e.g. the \(84\%\) quantile in \(\gamma\)  and the median
  in all other parameters for the corresponding bin) we can then generate a model using our Gaussian process emulator.
\end{itemize}
The result of this procedure is shown in \autoref{fig:gammacomp_28} for the power spectrum at redshift \(z=2.8\) which is the highest redshift showing a high \(\gamma\) value. We compare models generated in this way to the
measured power spectra shown as the blue and green points in the figure. Bands show the \(68\%\) confidence interval at each \(k\) for models generated using our emulator with random draws from the posterior distribution. Note that the forward model (due to both masking and forward modeling of noise and resolution) can generate slightly more converged model power spectra than the perfect model using the same parameters. The latter band is therefore actually a prediction for \(k\gtrsim \SI{0.02}{\skm}\) and its slightly larger extent is not surprising.
Also note that due to the way we chose to produce curves with different thermal parameters and the dimensionality of the space the range spanned by the dashed curves is typically smaller than the colored bands.
This is expected as the band shows the actual spread in the five/six (depending on the number of datasets used) dimensional parameter space whereas the lines are based on a quantile for one of the parameters and values at the center of the distribution close to that quantile for all others which will lead to a point inside the respective hypersurface, e.g. parameters of the purple/blue curve fall inside the five/six dimensional \(68\%\) surface, where the band corresponds to the actual surface).

We can see that all 5 models shown basically lead to the same power except for the
highest \(k\)-values measured \(k\ge 0.07\) (smallest scales).
At those scales a higher \(\gamma\) and lower \(\bar{F}\) indeed seems to provide a better fit to the data whereas at larger scales (smaller k) the model does not seem to be strongly affected by the parameters when moving along the degeneracy.

However, for other redshift bins (see \autoref{fig:gammacomp_all}) the sensitivity of the power spectrum toward changes in \(\gamma\) for a region around the median value shifts to different scales.
For example, at \(z\leq 2.0\) the most dominant effect seems to be on large scales, but note that we do not have the high precision BOSS measurement and that therefore both the range in allowed power spectra and the range of parameters in the \(2\sigma\) region of \(\gamma\) are larger.
All other redshifts seem to suggest a highest sensitivity to $\gamma$ at scales \(k\sim \SI{0.05}{\skm}\), different from both the lowest redshifts and \(z=2.8\).
While we note that differences between models of different \(\gamma\)
along the degeneracy direction are typically small compared to our
measurement errors for an individual \(k\)-bin, it is clear that the data of all bins combined has the precision to distinguish between these models, and that our inference is producing sensible fits. One might argue that the fact that the \(k\)-modes that are driving the fits
to high \(\gamma\) and low \(\bar{F}\) change for different redshift bins is a
source of concern, but we caution that the degeneracies in this
multi-dimensional parameter space are complex and not always easy to
visualize. We are confident that these results are not spurious, since
this high \(\gamma\), low \(\bar{F}\) combination persists consistently across
all redshift bins with \(z \le 2.8\), and both measurements and our inference
of different redshift bins are completely independent. We will return to this issue of discrepant \(\gamma\) and \(\bar{F}\) values in \autoref{sec:discussion} when we discuss possible systematic
errors in our hydrodynamical simulations.

\subsection{Measuring Thermal Evolution in the IGM using a Gaussian Prior on the Mean Transmission}
\begin{figure*}
\centering
\plotone{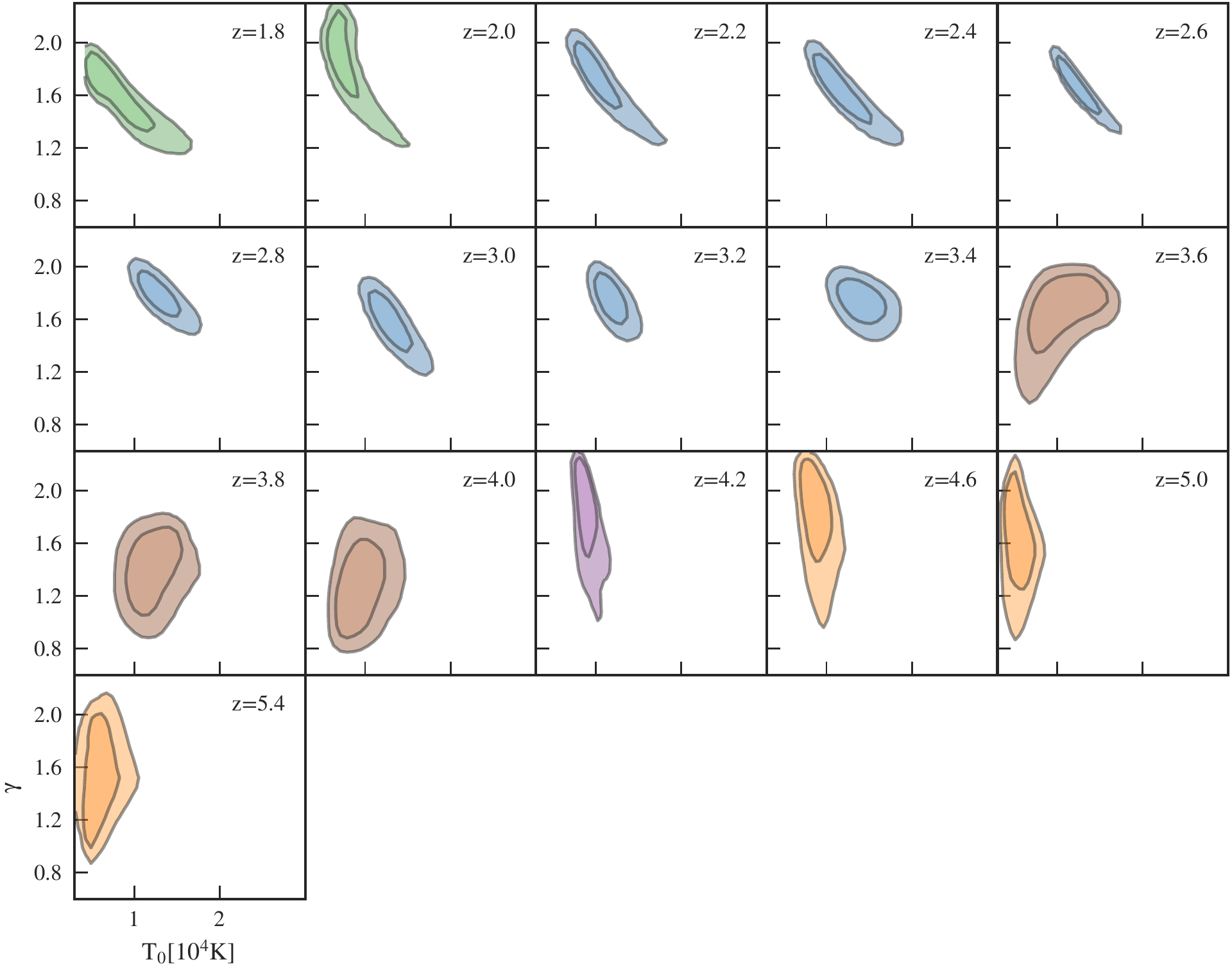}
\caption{Evolution of the \(T_0\) vs \(\gamma\) contours with redshift assuming the strong prior on \(\bar{F}\) for different combinations of datasets (different colors matched to \autoref{fig:evolution_onlydata}, filled contours showing our fiducial dataset while open contours show analyses ignoring the XQ-100 data as in the circles of \autoref{fig:evolution_onlydata}). When high resolution data is used we can see that strong constraints perpendicular to a degeneracy direction can be obtained. We can also see that this degeneracy direction rotates as the \ac{Lya} forest probes higher and higher densities.
}
\label{fig:evolution_2d_correlation_T_gamma}
\end{figure*}
\begin{figure*}
\centering
\plotone{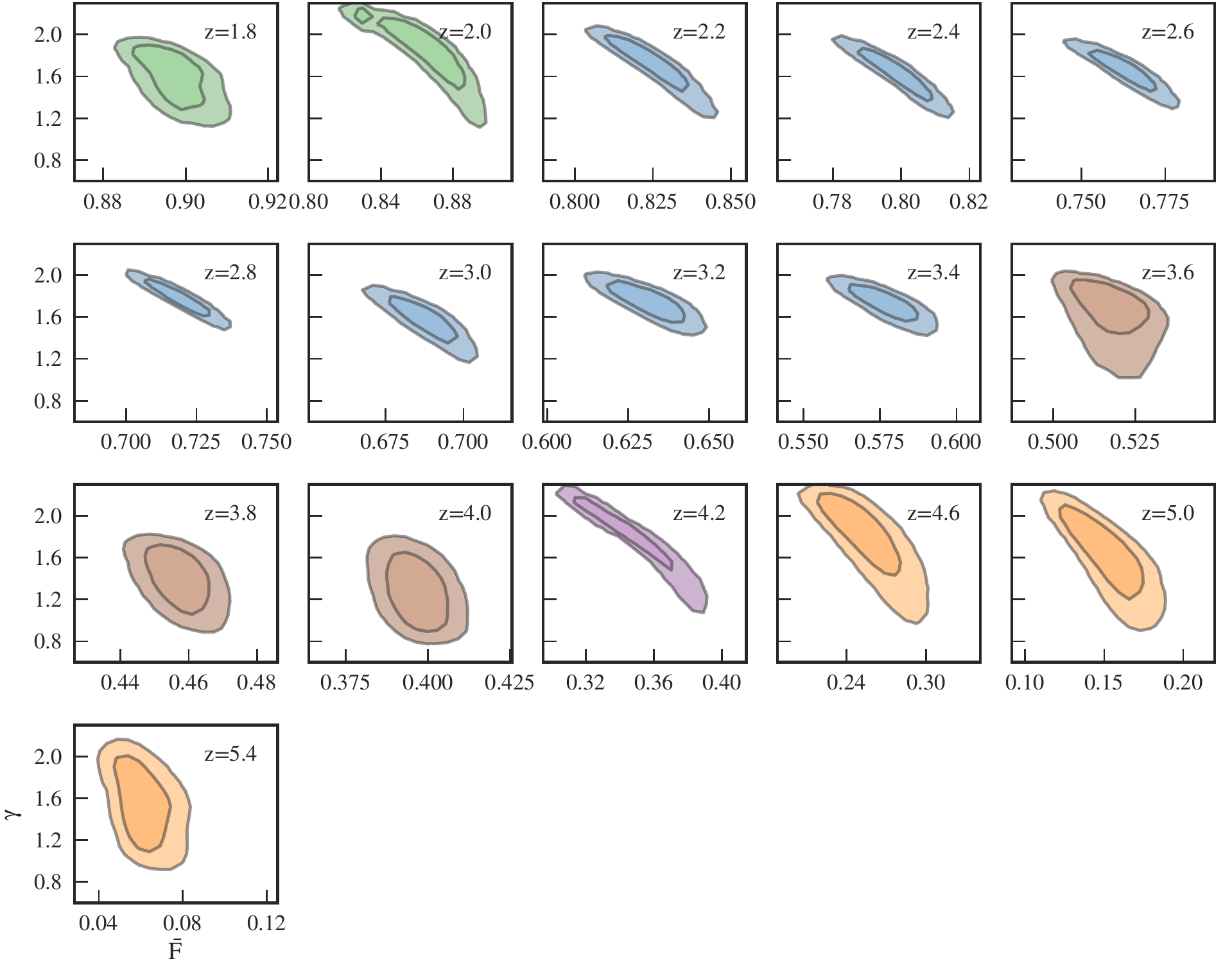}
\caption{The same as \autoref{fig:evolution_2d_correlation_T_gamma} but with \(\bar{F}\) vs. \(\gamma\) contours. This shows that independent of redshift \(\gamma\) and \(\bar{F}\) are strongly anticorrelated.
}
\label{fig:evolution_2d_correlation_Fbar_gamma}
\end{figure*}
\begin{figure*}
\centering
\plotone{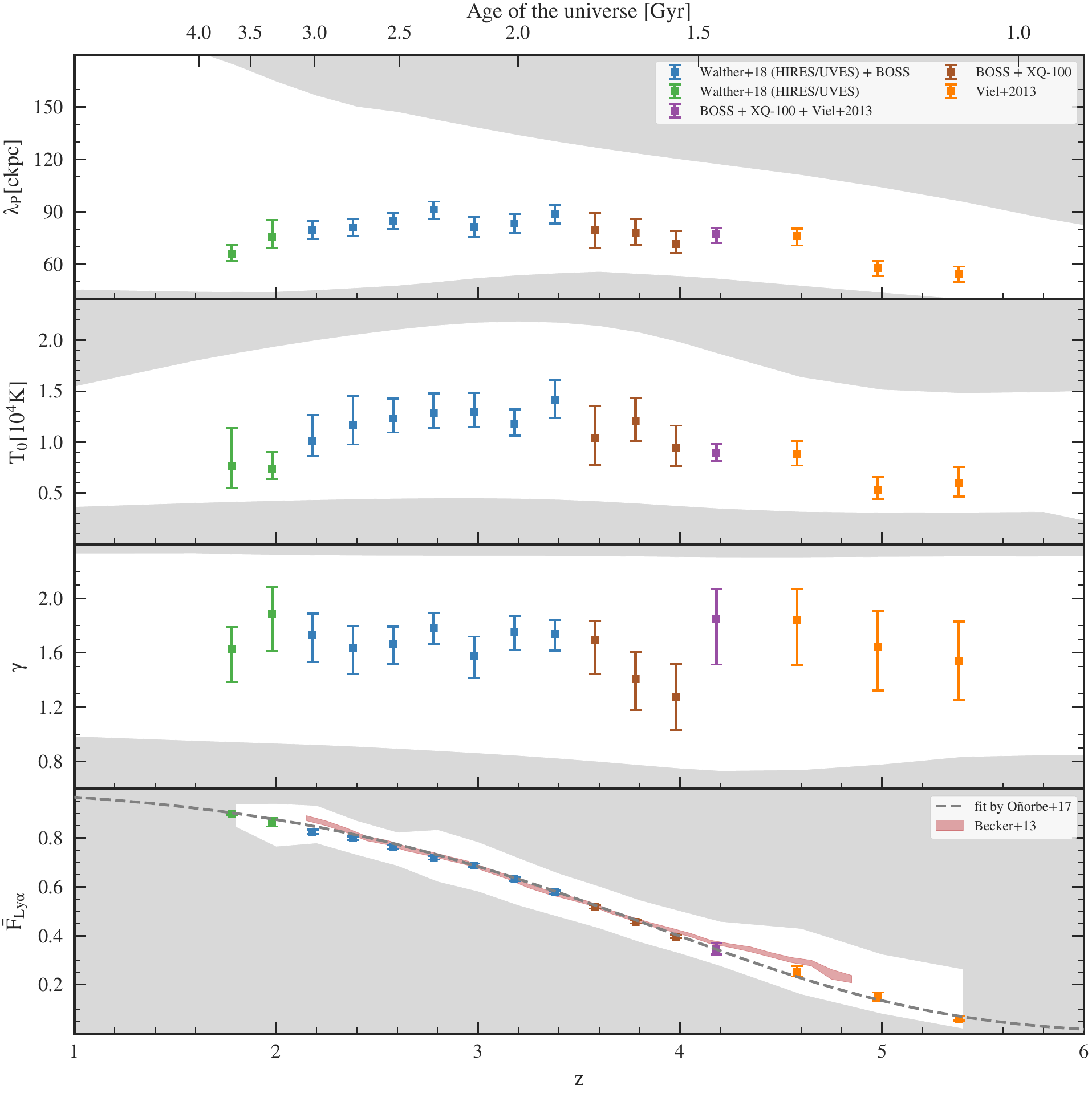}
\caption{The fiducial measurements from \autoref{fig:evolution_onlydata}, but assuming a Gaussian prior on the mean transmission complying with the fit of \citep{Onorbe2017ConstrainingReionization} measurement within errorbars given by observations \citep{becker2013refinedmeasurementmean,faucher-giguere2008EvolutionIntergalacticOpacity,kirkman2005HIopacityintergalactic}. We can see that now the obtained \(\gamma\) values at low redshifts are far lower (and compatible with the expected value of 1.6 long after reionization) due to the additional mean transmission constraint. The high values of \(\gamma\) at high redshifts obtained here, are likely due to the discrepancy of the mean of our chosen prior (dashed curve) with the \citet{becker2013refinedmeasurementmean} (red band) analysis for the mean transmission.
Due to the far lower overdensities probed at high-redshifts compared to low redshifts these high values of \(\gamma\) do not change results on \(T_0\) strongly as degeneracies are largely broken (see also the evolution of the \(T_0\)-\(\gamma\) and \(\gamma\)-\(\bar{F}\) contours which can be found in \autoref{fig:evolution_2d_correlation_T_gamma} and \autoref{fig:evolution_2d_correlation_Fbar_gamma}
).
}
\label{fig:evolution_onlydata_gaussian_1sig}
\end{figure*}
\begin{figure}
\centering
\plotone{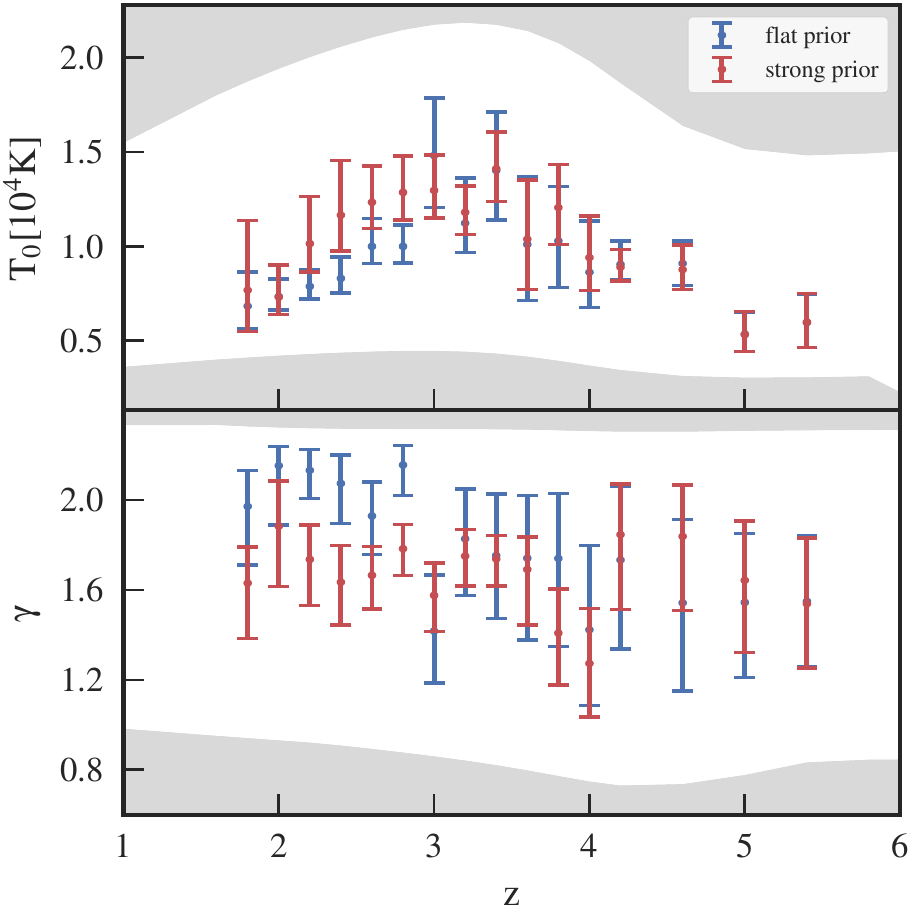}
\caption{Comparison of our results on the thermal state dependent on redshift between both priors (different colors). We can see that the flat prior leads to far higher values of \(\gamma\) that due to correlations between parameters lead to lower values of \(T_0\).
).
}
\label{fig:priorcompare}
\end{figure}

Given that independent precise constraints on the mean transmission exist we now consider the effect of applying a Gaussian prior on the mean transmission based on these measurements (see discussion in \autoref{sec:priors} for details). Henceforth we will refer to these fits as the `strong prior' results, and we will designate them as
our fiducial measurements (as opposed to the joint fits for thermal parameters and \(\bar{F}\) described in
previous sections). Note that most previous analyses of the IGM thermal properties have simply assumed
perfect knowledge of the mean transmission \citep[see][for exceptions]{lidz2010MeasurementSmallscale, Irsic2017Newconstraints}, such that this  `strong prior'
approach is more consistent with previous efforts.

We present the redshift evolution of posterior parameter degeneracies assuming the strong prior in \autoref{fig:evolution_2d_correlation_T_gamma}.
Each panel in these figures shows the 2d marginalized \(68\%\) and \(95\%\) confidence regions of \(T_0\) vs. \(\gamma\).
While \(\gamma\) and \(T_0\) are strongly anticorrelated at low redshifts \(z \leq 3.4\), i.e. the contours are close to diagonal, this correlation gets weaker at higher redshifts (especially at \(z\geq 4.2\)), i.e. contours become aligned with the axes due to lower overdensities probed by the power spectrum.
Likewise, the \(\gamma\) vs. \(\bar{F}\) confidence regions are shown in \autoref{fig:evolution_2d_correlation_Fbar_gamma}.
Note that these properties are correlated independent of redshift, in stark contrast to the thermal parameter degeneracy, while still changing shape and direction due to the different precision of the measurements.
Therefore, a change of prior for the mean transmission measurements propagates into \(\gamma\) at high redshifts (\(z\geq 4.2\)), but does not affect \(T_0\) significantly.
At lower redshifts (especially for \(z\leq 3.4\)), however, \(\gamma\) is strongly correlated with both \(T_0\) and \(\bar{F}\), so a change in priors for any of the three quantities always affects the results on the other two quantities as well.
Consequentially the change in our mean flux prior affects lower redshifts (especially \(z<3\)) more strongly than higher ones.

We show the fully marginalized posterior constraints on thermal parameters as a function of redshift in \autoref{fig:evolution_onlydata_gaussian_1sig}.
We can see that now the values of \(\gamma\) cover the theoretically expected value of \(\gamma\approx 1.6\) at low redshifts \(z<3\), while the values of \(T_0\) obtained are higher than in the fit using a flat prior on mean transmission because of the degeneracies between \(T_0\) and \(\gamma\).
This is more clearly illustrated in \autoref{fig:priorcompare} where we compare \(T_0\) and \(\gamma\) evolution for the different prior assumptions.
We can see that indeed the changes between both the two fits are strongly anticorrelated between \(T_0\) and \(\gamma\) and that the change in marginalized parameters between the two cases can be large, particularly for \(\gamma\) where the differences at \(2.2 \leq z \leq 2.6\) are $\gtrsim 2\sigma$ and as high as \(3\sigma\) at \(z=2.8\).

However, this is the first thermal evolution measurement performed over the whole epoch of \HeII{} reionization and beyond based on the power spectrum.
Using the strong mean flux prior we also obtained reasonable results including physically possible measurements of \(\gamma\), a rise in temperature for \(z\gtrsim 3\) as time progresses (or redshift decreases) and the first measurement of the IGM cooling down thereafter.
In the next sections we compare our strong prior results to recent thermal parameter measurements from different methods as well as models of IGM thermal evolution.

\subsection{Comparison to Previous Measurements}
\label{sec:previous_data}

\begin{figure*}
\centering
\plotone{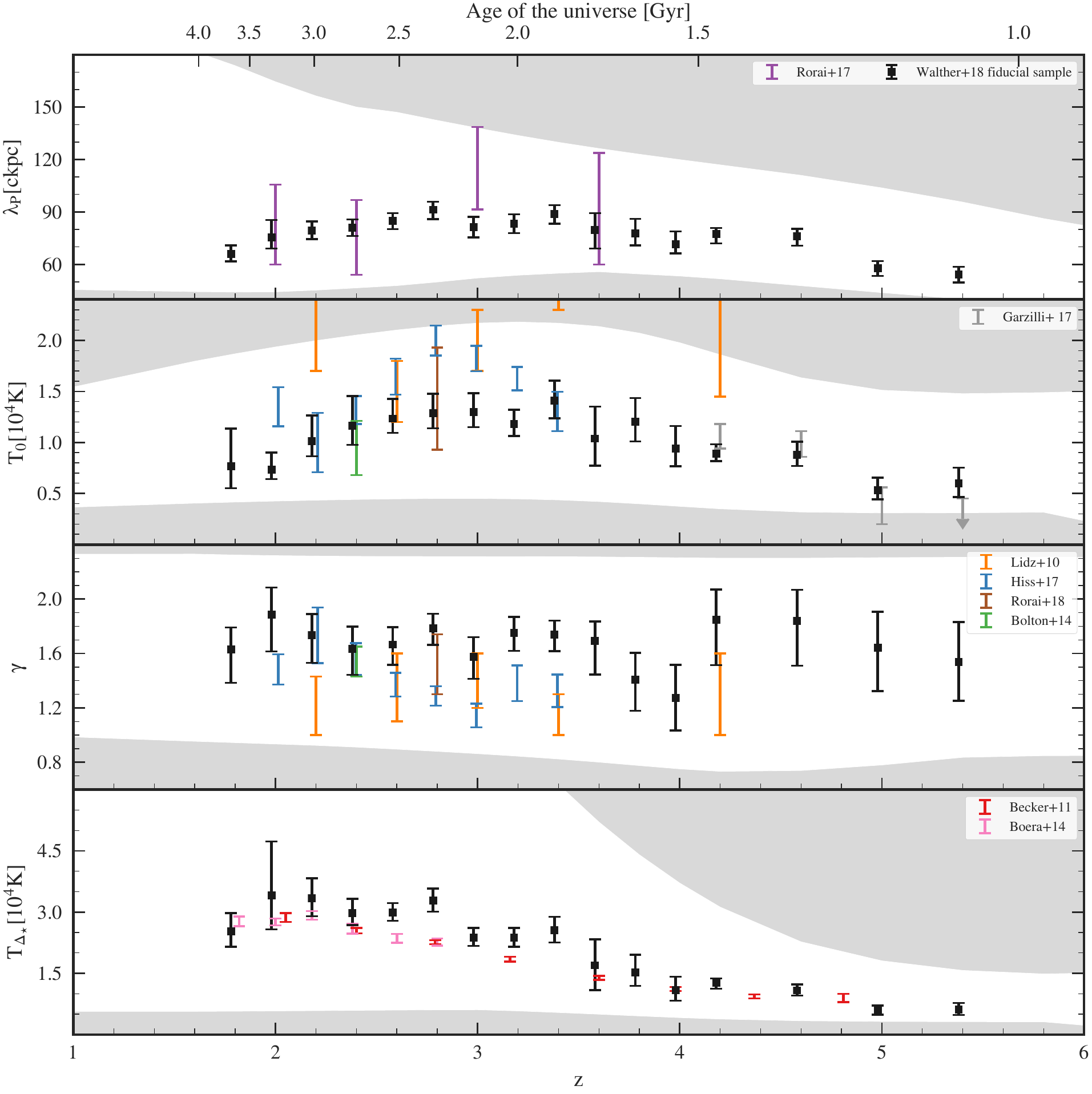}
\caption{The fiducial data from \autoref{fig:evolution_onlydata_gaussian_1sig} (black points) assuming the strong Gaussian prior on \(\bar{F}\).
In addition to the previous plots we show the thermal parameters as well as \(T(\Delta_\star)\) at the optimal overdensities \(\Delta_\star\) for curvature measurements as given by \citet{becker2011DetectionextendedHe}.
We compare to measurements of thermal evolution in the IGM based on different statistics: curvature (red, pink), line fitting (green, blue, brown), wavelets (orange), phase angles (purple) and power spectrum (gray).
We can see overall good agreement with previous datasets (except for wavelets) albeit significantly higher \(T(\Delta_\star)\) than in the curvature measurements is obtained at some redshifts.
All measurement errors shown are \(1\sigma\) or \(68\%\) intervals, for measurements that only quote \(2\sigma\) errorbars we divided those by a factor of two.}
\label{fig:compare_with_data_1sig_prior}
\end{figure*}

A comparison of our results to recent measurements of thermal parameters is shown in \autoref{fig:compare_with_data_1sig_prior}.
We  discuss the various datasets involved and elaborate on the comparison to our new measurement below.

The phase angle PDF of quasar pairs \citep{rorai2013NewMethodDirectly} measures the
smoothness of the 3d distribution of IGM gas and therefore directly
constrains the pressure smoothing scale \(\lambda_\mathrm{P}\) independent of the instantaneous thermal state of the IGM (i.e. \(T_0\) and \(\gamma\)).
\citet{Rorai2017Measurementsmall-scale} measured \(\lambda_\mathrm{P}\) from a sample of quasar pairs in 4 redshift bins between \(2.0 \leq z \leq 3.6\).
\autoref{fig:compare_with_data_1sig_prior} shows that our inferred values of \(\lambda_\mathrm{P}\) are fully consistent with the \citet{Rorai2017Measurementsmall-scale} measurement. We also see a smaller uncertainty in our power spectrum based measurement. Part of the explanation for these small error bars lies in how our model grid, and therefore our prior probability, is set up.
As discussed in \autoref{sec:priors}, the degeneracy between \(T_0\) and \(\lambda_\mathrm{P}\) within our simulated models combined with our approach of not extrapolating to regions outside our model grid result in a positive correlation in the prior probability between these parameters.
However, the power spectrum cutoff, is sensitive to a degenerate combination of both \(\lambda_\mathrm{P}\) and thermal broadening
\citep{Peeples2010Pressuresupporta,rorai2013NewMethodDirectly} leading to an anti-correlation in the likelihood.
So the correlations inside our prior (\autoref{fig:prior}) and the degeneracy direction of the likelihood due to the aforementioned effect are nearly perpendicular and as the posterior distribution is the product of these two, resulting constraints appear very tight. However, we argue that it is hard to generate physical models without imprinting the correlation between thermal state and \(\lambda_\mathrm{P}\) that depends on the integrated thermal history of the IGM.   While the uncertainties in \(\lambda_\mathrm{P}\) might still be somewhat underestimated, we note that our prior grid degeneracy has a strong physical motivation (see also \autoref{sec:priors}).

The orange points in \autoref{fig:compare_with_data_1sig_prior} show the \(T_0\) and \(\gamma\) measurements from
\citet{lidz2010MeasurementSmallscale}, who decomposed the \ac{Lya} forest of the \citet{dallaglio2008unbiasedmeasurementUV} dataset
into wavelets and analyze the PDF of their squared amplitudes to derive constraints on the thermal state of the IGM.
We note that this data is a subset of that used to compute the power spectrum in \citetalias{Walther2017NewPrecision} and analyzed here.  Note that their \(\gamma\) constraint is often limited to the boundaries of their fits\footnote{We therefore show the extent of their \(1\sigma\) contours (as a by-eye marginalization) for \(\gamma\) in the Figure.}.
While the wavelet analysis results at \(z=2.6\) are consistent with our measurement we disfavor the \(z=2.2\), \(z=3.0\), \(z=4\)  and especially \(z=3.4\) wavelet results which seem to indicate a far hotter IGM than our measurement. The origin of this discrepancy is unclear, but it was also noted before by \citet{becker2011DetectionextendedHe}.

Another method for obtaining constraints on the thermal state of the IGM is by decomposing the \ac{Lya} forest into individual absorption lines, assuming that a cutoff in the distribution of column densities \(N_\mathHI\) vs. Doppler parameter \(b\) exists and can be attributed to lines that are only thermally broadened \citep[see e.g.][]{schaye2000thermalhistoryintergalactic,rudie2012TemperatureDensityRelation,bolton2014consistentdeterminationtemperature,Hiss2017NewMeasurement,Rorai2018newmeasurement}.
Especially the new \citet{Hiss2017NewMeasurement} (which is based on the same dataset as \citetalias{Walther2017NewPrecision} and is using a subset of the same simulation grid) thermal evolution result seems to hint toward a period of heating until \(z\sim2.8\) that could be attributed to \HeII{} reionization.

For both \(T_0\) and \(\gamma\) we see broad agreement between our measurements and the line-fitting results at most redshifts.
Of particular interest are \(z=2.4\) and \(z=2.8\) where several line fitting measurements  exist.
At \(z=2.4\) we do reproduce the result from \citet{Hiss2017NewMeasurement} (blue points) as well as \citet{bolton2014consistentdeterminationtemperature} (green) in both \(T_0\) and \(\gamma\).
At \(z=2.8\) we agree with the \citet{Rorai2018newmeasurement} (brown point), but obtain higher precision.
However, agreement with \citet{Hiss2017NewMeasurement} at this redshift seems to be poor as they measure both higher \(T_0\) and lower \(\gamma\) (which is along the degeneracy direction for line fitting analyses as well as the power spectrum). Part of this discrepancy might come from systematics in the Voigt profile analysis depending on the cutoff fitting algorithm chosen, as \citet[][see Appendix B]{Hiss2017NewMeasurement} find either a multimodal posterior probability distribution for \(T_0,\gamma\) with a similar \(68\%\) confidence interval as the \citet{Rorai2018newmeasurement} or a unimodal distribution with the values shown here depending on the cutoff fitting algorithm used.
Whether this multimodal behavior results from systematics in the measurement procedure or is a real physical effect from e.g. a real multimodal IGM temperature density relation is not yet clear, but we do not see such behavior in our power spectrum analysis.
For the other overlapping redshifts (except \(z=2.2\) and \(z=2.4\) which match very well) we generally measure a lower \(T_0\) and higher \(\gamma\) compared to \citet{Hiss2017NewMeasurement}.

The most precise measurements of temperature in the IGM so far are based on the mean curvature in the \ac{Lya} forest \(\langle\kappa\rangle\) \citep{becker2011DetectionextendedHe,  boera2014thermalhistoryintergalactic}.
These measurements constrain \(T(\Delta_\star)\) at an optimal overdensity \(\Delta_\star\) at which a one-to-one relation between the mean curvature \(\langle\kappa\rangle\) of the \ac{Lya} forest and \(T(\Delta_\star)\) exists independent of the slope \(\gamma\) of the \ac{TDR} (note again \autoref{fig:evolution_2d_correlation_T_gamma} which shows the corresponding degeneracy for the power spectrum). However this method is not able to measure \(\gamma\) or $T_0$ independently.
To compare to curvature based measurements, we compute \(T_{\Delta_{\star}}=T_0 \Delta_{\star}^{\gamma-1}\) (using the values for \(\Delta_{\star}\) given by \citealt{becker2011DetectionextendedHe}) for each sample in our MCMC chain and evaluate the \(68\%\) confidence interval. This approach allows us to directly compare to what the curvature results measure.

The agreement with the curvature analysis seems to be generally good
for the largest part of the overlapping redshift range, but we seem to
measure overall slightly higher temperatures.  There are some
redshifts \(z=2.6, 2.8, 3.2, 3.4\) where our analysis gives
significantly higher temperatures than implied
by the curvature measurements. Note in particular that at \(z=2.8\) where
we see the strongest discrepancy between our results and the curvature measurements,
multiple measurements of the thermal state have been performed via several
different methods  and these results do not full agree with each other.
We argue, that the overall agreement is still good given the
significantly different datasets, statistical approaches and models
used for both types of analysis.
E.g., the difference in measured thermal state might potentially arise due to the different sensitivity of both statistics to metal contamination.
While the power spectrum is only weakly affected by residual metal lines on the very smallest scales we cover here \citep[see the comparison in][]{Walther2017NewPrecision}, the averaged squared curvature is basically measuring \(\int_{k_min}^\infty k^5 P(k) \mathrm{d}\ln k\) \citep[see Appendix D in][]{puchwein2015photoheatingintergalacticmedium} and thus enhances the weight of residual small scale contamination in the \ac{Lya} forest.
At the same time small scale contaminants, like e.g. leftover metal lines, would decrease the obtained IGM temperatures as there is now too much small-scale power, thus leading to a colder IGM in curvature than in power spectrum analyses.
Additionally, these measurements did not marginalize over the mean flux in the simulations, thereby e.g. potentially underestimate their errors.

Finally, we also show the \citet{Garzilli2017CutoffLyman-a} measurement of \(T_0\) at \(4.2 \leq z\leq 5.4\) based on the same \citet{viel2013Warmdarkmatter} dataset we use here (gray points, the limit is at the \(1\sigma\) level), but using a different analysis pipeline and including a \ac{WDM} particle mass as an additional free parameter.
We can see that for \(z\leq 5\) the agreement is good, but for \(z=5.4\) we seem to get slightly higher values of \(T_0\) than their \(1\sigma\) upper limit. Part of that difference can be attributed to the additional freedom in their model.

Overall we conclude, that the agreement between our data and previous results is reasonably good.
Our measurement comprises a strong advancement with regard to previous analyses especially due to the large range of uniformly covered redshifts and due to jointly constraining \(T_0\) and \(\gamma\) over this full range.

\subsection{Comparing to Thermal Evolution Models for Different \HeII{} Reionization Scenarios} \label{sec:model_comp}
\begin{figure*}
\centering
\plotone{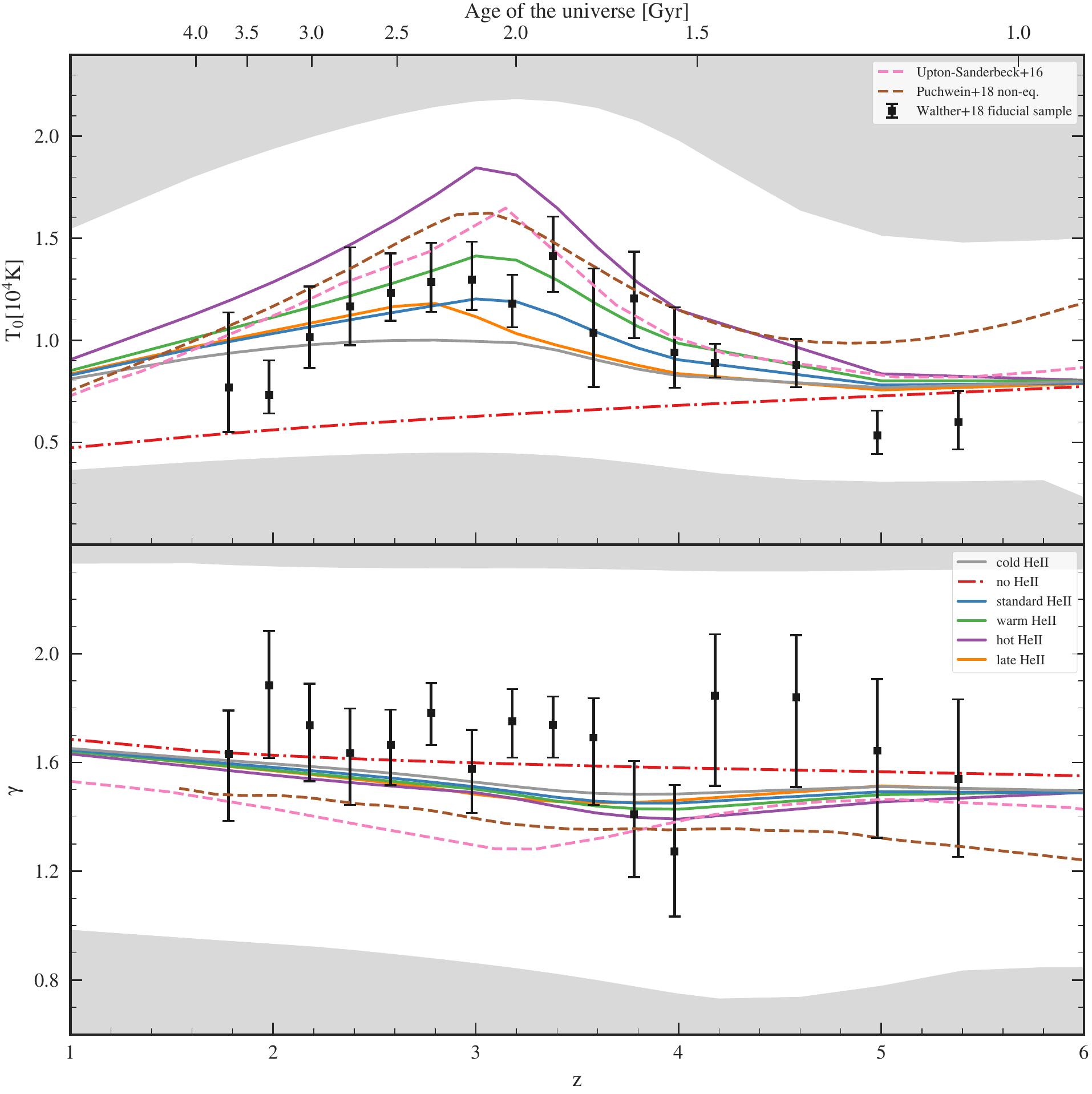}
\caption{
The fiducial data assuming the strong Gaussian prior from \autoref{fig:evolution_onlydata_gaussian_1sig} (black points) compared to thermal evolution models assuming different redshifts of \HeII{} reionization and heat inputs during this process (solid curves) and without any \HeII{} reionization (dot-dashed red curve). The model parameters are given in \autoref{tab:thermalpar_modelcomp}. We also show comparisons to the \citet{uptonsanderbeck2016Modelsthermalevolution} (dashed pink) thermal evolution model and a run using the \citet{Puchwein2018Consistentmodelling} non-eq. heating rates in a Nyx simulation (dashed brown).
We can clearly see that the data shows a hotter IGM than created in the model without \HeII{} reionization.
Instead, the overall evolution of thermal parameters seems to agree well with the standard to warm \HeII{} reionization scenarios in both \(T_0\) and \(\gamma\).
Finally, the temperatures found at the 2 highest redshift bins are colder than any model.
}
\label{fig:compare_with_models_1sig}
\end{figure*}
\input{modeltable}
In the previous sections we performed a self-consistent measurement of thermal evolution in the IGM from \(z=5.4\) to \(z=1.8\) corresponding to \(\SI{3}{\giga\yr}\) of cosmic history.
In this section we compare to simulations to thermal evolution due to \HeII{} reionization as this is expected to be the dominant process setting the thermal state of the IGM at this epoch.

In \autoref{fig:compare_with_models_1sig} we show comparisons between our thermal evolution measurement and models based on different approaches. The solid curves show the "explicit reionization" simulations from our model grid
for which hydrogen reionizes (to a level \(x_{\mathHII}=99.9\%\), note that for our models this point is typically reached with a delay of \(\Delta z\approx 1\) compared to the corresponding \(z_\mathrm{reion, 50}\)\footnote{Notice that the interpretation of the duration of reionization in homogeneous UVB models can be misleading. We point to \citep{Onorbe2017Self-consistentModeling} for a full discussion in this regard.}) at \(z_\mathrm{reion,\mathHI} =6.5\) in agreement with the \citet{PlanckCollaboration2016Planckintermediate} (and also \citealt{PlanckCollaboration2018Planck2018}) constraints, but for which the parameters governing  \HeII{} reionization are varied (see \autoref{tab:thermalpar_modelcomp}).
The red dash-dotted curve is showing an extreme version of these models for which  \HeII{} was never reionized\footnote{We note that this reionizes \HI{} slightly earlier \(z =7.3\) which is still in good agreement with both Planck results.}.
The measured temperature at mean density is significantly higher (by \(\gtrsim 3\sigma\) for each of the 7 individual redshift bins with \(2.2\leq z\leq 3.4\)) than this "no \HeII" scenario for \(z\leq4.6\) suggestive of a period of \HeII{} reionization taking place.

To allow a comparison between different \HeII{} reionization scenarios, the gray, blue, green, and purple curves show models with \(z_\mathrm{reion,\mathHeII}=3.0\) assuming different amounts of heat being injected \(\Delta T_{\mathHeII}\) into the IGM varying from \(\SI{10000}{\kelvin}\) (cold) to \(\SI{30000} {\kelvin}\) (hot); whereas the orange curve shows a model with \(\Delta T_\mathHeII=\SI{15000}{\kelvin}\) but \(z_\mathrm{reion,\mathHeII}=2.8\) (late).
We can see that the models predict an extended period of heating (i.e. increasing \(T_0\)) until \(z_\mathrm{reion,\mathHeII}\) followed by the IGM cooling down due to the expansion of the universe whose effects on the thermal state cannot be fully counteracted by ionizations anymore.
Overall, for our measurement this rise and fall in \(T_0\) lies between the standard and warm \HeII{} evolution models for \(2.2\leq z\leq 4.6\) disfavoring particularly hot or late phases of \HeII{} reionization.

We also compare to the analytical thermal evolution model by \citet{uptonsanderbeck2016Modelsthermalevolution} and the fiducial non-equilibrium reionization model by \citet[][see their Figure 6]{Puchwein2018Consistentmodelling}.
We note that while the general shape of thermal evolution looks similar to both models for \(z\leq 4.6\) we seem to obtain a slightly less pronounced peak in \(T_0\).
Overall, the temperature evolution we see in this redshift range is indeed well modeled by an \HeII{} reionization event followed by photoionization equilibrium in an adiabatically expanding IGM.
While it has been argued that this effect has been seen before \citep{becker2011DetectionextendedHe}, previous work did not break the degeneracy between \(\gamma\) and \(T_0\).
Note that also the cooldown of the IGM after reionization has never been conclusively observed due to this degeneracy.

Models of \HeII{} reionization typically also show a dip in \(\gamma\) resulting from the IGM to be more isothermal during reionization events \citep[see e.g. also][]{mcquinn2009HeIIReionization}. We can also see this effect by comparing \(\gamma\) for our \HeII{} models with the no-\HeII{} model.
Note that while the \citet{uptonsanderbeck2016Modelsthermalevolution} model also shows a dip (albeit at later times and with a more strongly isothermal \(\gamma\)), the fully non-equilibrium simulation by \citet{Puchwein2018Consistentmodelling} does show an intrinsically smaller \(\gamma\) and no strong dip. The reason for this is that the non-equilibrium model reached \(\gamma=1\) at \(z=7\) due to \HI{} reionization and is still recovering from this feature, i.e. it did not yet forget about the timing of \HI{} reionization. The "dip" for this model therefore manifests in the near constant evolution from \(z\sim 5\) to \(z\sim 3\) compared to an otherwise expected rise in \(\gamma\).

We can see this dip in \(\gamma\) for the measurement at \(z\sim 3.9\) aligned in redshift with the expected decrease due to \HeII{} reionization in
our explicit \HeII{} reionization models.
Note that the dip is only \(\sim 2 \sigma\) significant compared to the no-\HeII{} reion model, but overall a slightly higher value for \(\gamma\) than this model is preferred.
Also note that on the data side this feature is currently dominated by XQ-100 data (which is the highest resolution data available at \(3.6 \leq z\leq 4.0\)) which we strongly degraded by marginalizing over resolution.
Additional high resolution data or an accurate determination of the XQ-100 dataset resolution at these redshifts and
adopting a prior based on those results could therefore lead to additional constraints on \HeII{} reionization due to its signature in the slope of the \ac{TDR}.

Note that this feature also strongly relies on precise knowledge of \(\bar{F}\) as the expected decrease is very shallow. Additionally, \(\gamma\) values for \(z>4.2\) might have a significant uncertainty as measurements of the mean transmission get less accurate for this range due to the smaller amounts of data available and stronger fluctuations in the ionization state of the IGM.
Thus, there are currently several discrepant measurements of \(\bar{F}\) (as discussed in \autoref{sec:priors}) which consequently lead to a high uncertainty in \(\gamma\).

At early times (\(z \geq 5\), we call those points the highest redshift measurements)  we can see that the measured \(T_0\) is lower than in any of the models. Note again that similarly low temperatures were also obtained by \citealt{Garzilli2017CutoffLyman-a} based on the same dataset in a fully independent analysis.
While one could in principle think that an earlier redshift of \HI{} reionization gives the IGM more time to cool thereafter leading to lower temperatures at these times,
models suggest that is not the case and \(T_0\) has essentially forgotten about the timing of reionization by \(z=5.4\) (see e.g. \citealt{Onorbe2017ConstrainingReionization} who present models for a range of different \(6.0<z_\mathrm{reion,\mathHI}<9.7\)).
Instead the post-reionization thermal state mostly depends on the spectral shape of the \ac{UVB} \citep{McQuinn2016intergalactictemperature-density} and a low temperature at \(z\sim 5.4\) requires lower photoheating rates, i.e. a particularly soft \ac{SED} for ionizing sources is needed which is not favored by current models of the UVB \citep{Faucher-Giguere2009NewCalculation,haardt2012RadiativeTransferClumpy,Stanway2016Stellarpopulation,DAloisio2018HeatingIntergalactic,Puchwein2018Consistentmodelling}.

While it may be that the thermal state at $z \simeq 5.4$ would still be sensitive to the reionization redshift for particularly late \(z\lesssim 6\) reionization scenario \citep[which now seems to be allowed regarding the newest CMB results from][]{PlanckCollaboration2018Planck2018}, this would nevertheless need to be in conjunction with very low
reionization heat injection.  The recent results by
\citet{DAloisio2018HeatingIntergalactic} who use radiative transfer to
simulate photoheating by ionization fronts during \HI{} reionization
suggest that such low levels of IGM heating are unlikely.
Finally, note that the onset of \HeII{} reionization can only increase model temperatures and therefore worsen the disagreement as none of the models shown exhibits any \HeII{} reionization before \(z=4.8\).

Therefore, the small temperatures we (and also other groups using the same dataset) obtain at \(z\geq 5\) are challenging to fit with current models of reionization.
Consequently, models fitting the low-T measurements would also lead to a colder IGM at later times without additionally increasing heating due to e.g. \HeII{} reionization.
However, as current constraints at the highest redshifts rest upon the single dataset by \citet{viel2013Warmdarkmatter} based on a handful of objects, future measurements based on larger samples of quasar spectra obtained might change those low-\(T_0\) results.

\section{Systematic Effects on the Measured Thermal Evolution}\label{sec:discussion}
\begin{figure*}
\centering
\plotone{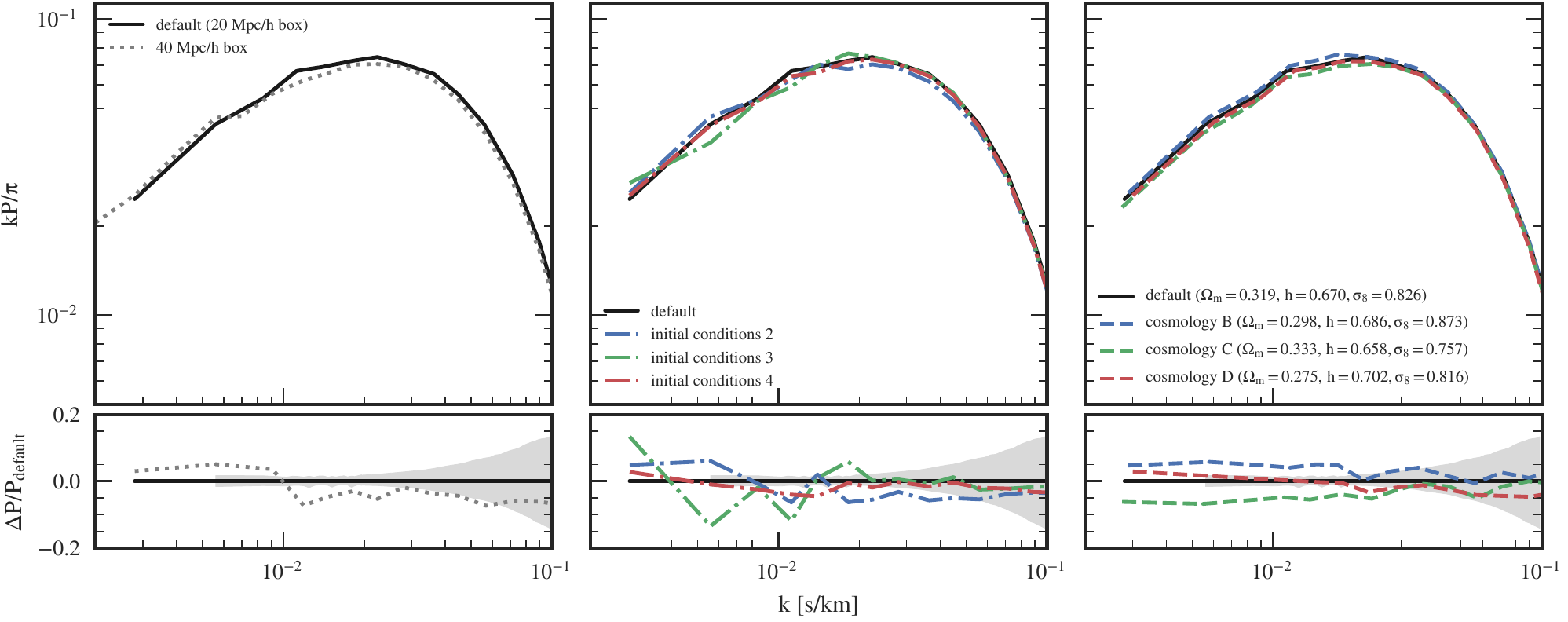}
\caption{Left: One model from our thermal grid at \(z=2.8\) (solid black, the redshift is taken as an example, other redshifts are similar) compared to the same model run with a two times larger box and the same spatial resolution (dotted gray). The bottom panel shows relative differences and the size of the \(68\%\) confidence region of jointly fitting BOSS + high-resolution data as a grey band.
Center: A comparison between different initial conditions (dot-dashed) that were elsewise run with the same setup.
Right: A comparison of models based on other cosmologies (B,C are compatible with the \citealt{PlanckCollaboration2016Planck2015} parameters and chosen to maximally change the matter power spectrum w.r.t. the default model, see \citealt{Onorbe2017Self-consistentModeling} for details; D is the cosmology from \citealt{lukic2015Lymanforestoptically}).
We can see that all three effects change the power on the \(5\%\) level.
}
\label{fig:model_effects}
\end{figure*}
\begin{figure}
\centering
\plotone{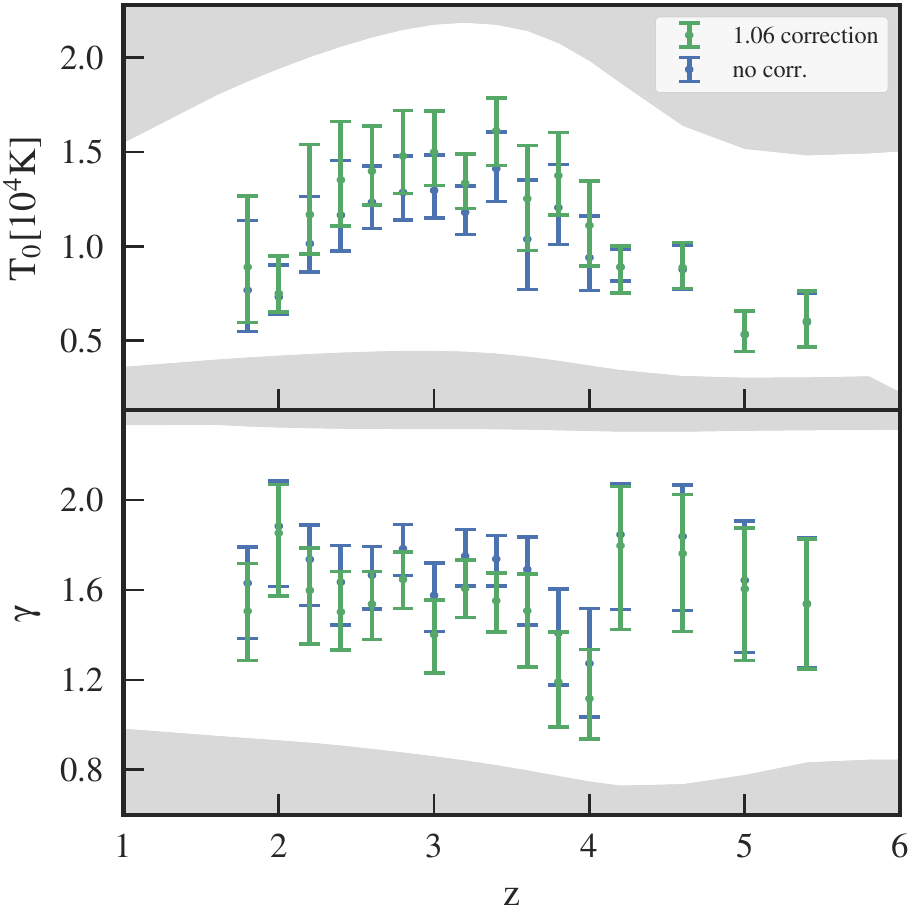}
\caption{Comparison of our results on the thermal state with fits obtained if we apply a flat "correction factor" of \(0.94\) (mimicking the joint effect of box size and cosmic variance seen in \autoref{fig:model_effects}) to the model power (different colors). We can see that without the "correction" higher values of \(\gamma\) as well as lower values of \(T_0\) (due to correlations between parameters) are obtained.
}
\label{fig:rescalecompare}
\end{figure}

In \autoref{sec:measurements-degeneracies} we attempted to jointly fit the mean flux and thermal parameters and arrived at puzzling results for \(\gamma\).
This, combined with the fact that independent high-precision measurements of the mean flux are available and that the most precise former analyses of thermal evolution fixed the mean flux, led us to adopt the strong prior which led to sensible results on thermal evolution of the IGM that are in broad agreement with previous measurements as well as simulation predictions.
In this section we investigate possible systematics in our modeling procedure which could be responsible for the high \(\gamma\)- low \(\bar{F}\) we observe with the flat prior on \(\bar{F}\) in \autoref{sec:results}.

We think that the biggest issue is our modeling and there are several possible sources of bias for our measurement: the small boxes used and the cosmic variance, not simultaneously exploring cosmological parameters, and spatial resolution of the simulation.
We attempt to quantify the significance of all these issues below. While ideally a large set of simulations would be used to do a detailed study of each issue, due to computational cost we are limited to a handful of simulations per problem.

To explore box size we compare one model from our grid to a simulation with exactly the same thermal model and cosmology performed with the same resolution, but with twice the box size, i.e. \(L_\mathrm{box}=\SI{40}{\hinvMpc},\) \(N_{cell}=2048^3\)\footnote{Note that the initial conditions cannot be the same for two boxes of different size and so every comparison of this kind includes cosmic variance on both boxes,but with \(\sqrt{8}\) times lower amplitude at a given mode for the larger box.}
In the left panel of  \autoref{fig:model_effects} we show this comparison.
Similar to the results of \citet{lukic2015Lymanforestoptically}, one clearly sees that for the range of power spectrum modes that we fit a \(\sim 6\%\) bias in the power might be expected due to box size effects. The gray curve shows the posterior \(68\%\) model interval from \autoref{fig:power_fit_singlez} as a measure of the joint precision of all datasets used in the fit.
So especially for scales \(k\lesssim\SI{3e-2}{\skm}\) box size effects are larger than this precision and could thus strongly affect the results.
Whether the overall \(6\%\) at \(k\gtrsim \SI{0.01}{\skm}\) results from box size effects or cosmic variance (see below) is unclear, but assuming the former, we perform an estimate of how much a flat bias affects our thermal evolution constraints.
For this purpose, we repeat our data analysis, but rescale the emulated power spectrum for every redshift by a factor of \(0.94\) independent of \(k\) and model parameters.
In \autoref{fig:rescalecompare} we show our fiducial analysis (blue) compared to this "corrected" measurement (green).
We can see that the rescaling leads to a \(\sim 0.5\sigma\) to \(1.2\sigma\) higher \(T_0\) and lower \(\gamma\) for all \(2.2\leq z \leq 4\).
Therefore, our measurement is clearly limited by the combined effect of box size and cosmic variance in this redshift range.
Note that the change when applying this rescaling is such that the inferred \(\gamma\) is reduced, i.e. the discrepancies we analyzed in \autoref{sec:discrepancies} become weaker.

Simulations also suffer from statistical variance for the largest modes
where the sampling is poor. To better understand this issue we ran
simulations with different initial conditions but an otherwise
identical setup.  The comparison of those runs to our default
simulation is shown in the middle panel of
\autoref{fig:rescalecompare}. We can clearly see, that even with just
4 samples of initial conditions a \(\sim 5\%\) change in the power can
be reached on small scales similar to the results in the boxsize test
above. Additionally, the effect of cosmic variance on the largest
scales (lowest \(k\lesssim \SI{0.01}{\skm}\)) can exceed the \(10\%\)
level, which is huge compared to the \(\sim 2\%\) errors of the BOSS
measurement.  To get both box size and cosmic variance effects under
better control requires an analysis based on larger simulations, where
doubling the (linear) box size would be expected to reduce cosmic
variance by a factor of \(\sqrt{8}\) (but also needs at least eight
times more computing time).

To understand the effect of cosmological parameters on the \ac{Lya} forest power spectrum we
compare to three different cosmologies consistent with the \citet{PlanckCollaboration2016Planck2015} results.
Cosmology B \& C were selected from their posterior distribution  in order to differ as much as possible in the linear matter power spectrum \citep[see][]{Onorbe2017Self-consistentModeling}.
Cosmology D uses the same parameters as in \citet{lukic2015Lymanforestoptically}.
The right panel of \autoref{fig:rescalecompare} shows that a change in cosmological parameters within the current CMB constraints can lead to a \(\sim 5\%\) change in the flux power as well.
Of course a more detailed analysis of this effect is needed and ideally one would marginalize over cosmological parameters adding additional dimensions to our simulation grid. However future independent higher precision cosmological constraints from either joining existing datasets or new measurements will reduce the strength of this effect.

Finally, the finite resolution of the simulations is not an issue at \(z\lesssim 4\) \citep[see Figure 11 in][showing convergence to \(1\%\) at \(z\leq 3\) and to better than \(5 \%\) at \(z=4\)]{lukic2015Lymanforestoptically}, but might be of some importance at \(z\gtrsim 5\) \citep[see Appendix of][]{Onorbe2017Self-consistentModeling} and might be more severe in exceptionally cold models as pressure smoothing is then weaker and structures are thus harder to resolve.
In the latter case the power at the smallest mode covered in our analysis could be underestimated at the \(\sim 10\%\) level which is comparable
to its errorbars. However, in contrast to box size effects only the smallest scales (\(k\gtrsim \SI{0.07}{\skm}\)) are affected which will not lead to changes as dramatic as seen for the other modeling errors considered in this section. However, the scale dependence of this effect, large scales (small \(k\)) being nearly unaffected while small scale power is reduced in the model, might lead to slightly underestimated results on \(T_0\).

In summary, we have seen that all four effects we discuss in this section, box size, initial conditions, cosmology, and resolution can affect the power spectrum by a similar amount as our statistical measurement errors at least for some range of scales and redshifts. We have seen that these effects can be comparable or larger than our statistical errors on
the power spectrum, and
can thus systematically change our thermal evolution at the \(0.5\) to \(1\sigma\) level at a range of redshifts.
Note again that all the effects we considered here are converged at the \(\sim 5\%\) level and a better treatment of any of the effects would require additional computation time or reduce the number of simulations that can be performed thereby increasing interpolation errors.
The current analysis is therefore the best compromise between accurate results and available computing time.
But note that the effects discussed here, might very well explain some of the discrepancies between constraints of the thermal state obtained by different groups.

\section{Conclusions and Outlook}\label{sec:conclusions}

In this work, we presented the first uniform thermal evolution
analysis based on the \ac{Lya} forest power spectrum covering a large
redshift range from \(z=5.4\) to \(z=1.8\) or equivalently a timespan
of nearly \(\SI{3}{\giga\yr}\).  For this purpose we combined multiple
high-precision measurements performed by several groups using
different instruments.  Furthermore, we compare this dataset with a
large grid of high-resolution hydrodynamical simulations to connect
the measured \ac{Lya} forest to physical properties of the IGM.  To
interpolate between these simulations we developed a Gaussian process
emulation scheme and take its errors into account using a
cross-validation approach.  Compared to previous results we measure
thermal evolution from high redshifts \(z=5.4\) to the limit of
\ac{Lya} forest observability with ground based telescopes due to the
atmospheric UV cutoff at \(z\sim 1.8\), and our combination of
high-precision low-$k$ measurements with our new high-$k$ analysis
allows us to break the well known degeneracy between the temperature
at mean density and the slope of the \ac{TDR}.
Our analysis thus provides the first comprehensive homogeneous analysis of IGM thermal evolution probing times as early the end stages of
\HI{} reionization, extending through the epoch of \HeII{} reionization, and spanning  the era of galaxy formation.

Our primary results are measurements of \(T_0\), \(\gamma\), and \(\lambda_P\) (see \autoref{tab:results_strong}) marginalizing over the mean transmission in two different ways (with a flat prior or a Gaussian prior based on recent measurements).
These measurements show a clear increase in \(T_0\) from \(T_0\sim\SI{6000}{\kelvin}\) at \(z=5.4\) to \(T_0\sim\SI{14000}{\kelvin}\) at \(z=3.4\) followed by a decrease reaching \(T_0\sim\SI{7000}{\kelvin}\) at \(z=1.8\).
We compared our results to published thermal evolution constraints using different statistics and find broad consistency with data from curvature, Voigt profile fitting and the phase angle distribution analyses.
Comparing  to simulations we indeed see compelling evidence for \HeII{} reionization in the rise of \(T_0\) which is not expected in absence of \HeII{} reionization.
In general the thermal parameters we obtained from fitting the power spectrum measurements agree well with models for which \HeII{} reionization is complete at \(z\sim 3\).
At later times, i.e. \(z<3\), we see the first conclusive evidence that the IGM is cooling down after the last reionization heating episode driven by
adiabatic cooling due to the expansion of the universe.

However, at the highest redshifts \(z\geq 5\) we find evidence for low temperatures \(T_0\sim \SI{6000}{\kelvin}\) (slightly higher, but consistent with other measurements based on the same dataset) that might be hard to explain with our current understanding of the shape of the \ac{UVB} at those redshifts as well as our current understanding of \HI{} reionization.
This is especially important as the same dataset resulting in these low temperatures also places the most stringent limits on the mass of \ac{WDM} \citep{viel2013Warmdarkmatter}. Comparing these power spectrum measurements to models that include both WDM particle mass as well as the IGMs thermal history as free parameters would necessarily result in an even colder IGM, because small-scale structure in the Ly$\alpha$ forest can now be erased by both thermal broadening and a finite WDM free-streaming length (see \autoref{fig:compare_with_data_1sig_prior}, compare to \citealt{Garzilli2017CutoffLyman-a}).
Thus,
given our current expectations for reionization heating, the
cold temperatures we infer provide additional evidence for a cold
dark matter universe.

To obtain a complete measurement of the \ac{IGM}'s thermal state, \ac{Lya} forest measurements clearly need to be extended to both higher and lower redshifts.
At high-\(z\) this would allow for testing the current power spectrum results and enable stronger joint constraints on the thermal state just after \HI{} reionization as well as the nature of dark matter \citep[see][for a forecast of possible constraints using high-resolution data up to \(z=6\)]{Onorbe2017ConstrainingReionization} due to an increase in the available dataset size in recent years.
At the same time, the great success of the COS and STIS instruments on HST enables new measurements of the thermal state at low redshifts (\(z\lesssim 1\)) allowing to test if the \ac{IGM} cools down further as theoretically expected. As the post-reionization \ac{IGM} physics is in principle well understood, these low-redshift measurements could then be used to constrain heat input from other astrophysical processes, e.g. galaxy formation or blazar heating.

However, to get precision constraints of the thermal state in the IGM better hydrodynamical simulations are needed.
We characterized the effect of box size, cosmic variance and cosmology and found, that for some range of scales systematic uncertainties due to these effects can be comparable to our measurement precision.
Future progress will therefore rely on simulating larger grids to marginalize over cosmological parameters or alternatively a more precise external determination of those parameters as well as larger simulation boxes.
Thanks to great improvements in recent years, allowing nearly linear scaling of computing time with volume (at fixed resolution) in some hydrodynamical simulation codes, and the current advancement of computing speed in supercomputers this will be possible within the next few years.

\acknowledgments
We thank Ashmeet Singh \& Dan Foreman-Mackey for helpful discussions about Gaussian processes.

We thank Hector Hiss \& Vikram Khaire for comments on the manuscript as well as all members of the ENIGMA group\footnote{\url{http://enigma.physics.ucsb.edu/}} at the Max Planck Institute for Astronomy (MPIA) and University of California Santa Barbara (UCSB) for insightful suggestions  and  discussions.

Calculations presented in this paper used the hydra and draco clusters clusters of the Max Planck Computing and Data Facility (MPCDF, formerly known as RZG). MPCDF is a competence center of the Max Planck Society located in Garching (Germany).
We also used resources of the National Energy Research Scientific Computing Center (NERSC), which is supported by the Office of Science of the U.S. Department of Energy under Contract No. DE-AC02-05CH11231.

ZL was in part supported by the Scientific Discovery through Advanced Computing (SciDAC) program funded by U.S. Department of Energy, Office of Advanced Scientific Computing Research and the Office of High Energy Physics.

\facilities{
VLT/ UVES \citep{dekker2000Designconstructionperformance},
Keck/ HIRES  \citep{vogt1994HIREShighresolutionechelle},
SDSS-III/ BOSS \citep{Dawson2013BaryonOscillation},
VLT/ XSHOOTER \citep{Vernet2011X-shooternew},
Magellan/ MIKE \citep{Bernstein2003MIKE:Double}}
\software{emcee \citep{foreman-mackey2013emceeMCMChammer}, george \citep{ambikasaran2016FastDirectMethods}, numpy \citep{vanderWalt2011NumPyArray:}, scipy \citep{Oliphant2007PythonScientific}, astropy \citep{AstropyCollaboration2013Astropy:community}, matplotlib \citep{Hunter2007Matplotlib:2D}, seaborn \citep{Waskom2017mwaskom/seaborn:v0.8.1}, Nyx \citep{almgren2013NyxMassivelyParallel}}

\bibliography{thermal}

\appendix

\section{Tables of the Measured Thermal Evolutions} \label{sec:tables}
\input{MCMC_results_flat}
\input{MCMC_results_strong}
In this section we tabulate our measurement values at each redshift for the flat prior on \(\bar{F}\) (\autoref{tab:results_flat}) and the strong prior (\autoref{tab:results_strong}). Those tables do not only show the marginalized constraints of all thermal parameters, but additionally show values for the temperature at the overdensity \(\Delta_\star\) where curvature measurements are optimal \citep[with the value for \(\Delta_\star\) interpolated in redshift between results from][]{becker2011DetectionextendedHe} as well as at \(\Delta_\mathrm{power}\) where the degeneracy between \(\gamma\) and \(T\) is minimized for the power spectrum. The latter was obtained by assuming a power law relation \(T(\Delta_\mathrm{power})=T_0 \Delta_\mathrm{power}^{\gamma - 1}\) to the samples in our Markov chains and varying \(\Delta_\mathrm{power}\) such that the variance of \(T(\Delta_\mathrm{power})\) is minimized. The density values where degeneracies are minimal are tabulated as well.
We will provide chains from our MCMC analysis on request.

\begin{acronym}
 \acro{DLA}{damped \acs{Lya} absorption system}
 \acro{UVB}{ultraviolet background}
 \acro{LLS}{Lyman Limit System}
 \acro{pLLS}{partial \ac{LLS}}
 \acro{Lya}[Lyα]{Lyman Alpha}
 \acro{DM}{dark matter}
 \acro{UVES}{Ultraviolet and Visual Echelle Spectrograph}
 \acro{HIRES}{High Resolution Echelle Spectrometer}
 \acro{KODIAQ}{Keck Observatory Database of Ionized Absorption toward Quasars}
 \acro{DR}{Data Release}
 \acro{VLT}{Very Large Telescope}
 \acro{SDSS}{Sloan Digital Sky Survey}
 \acro{BOSS}{Baryon Oscillation Spectroscopic Survey}
 \acro{IGM}{intergalactic medium}
 \acro{MCMC}{Markov Chain Monte Carlo}
 \acro{1DPS}{1d flux power spectrum}
 \acro{FGPA}{fluctuating Gunn-Peterson absorption}
 \acro{PDF}{probability density function}
 \acro{PCA}{principal component analysis}
 \acro{AMR}{adaptive mesh refinement}
 \acro{PM}{particle-mesh}
 \acro{BAL}{broad absorption line}
 \acro{EoR}{Epoch of Reionization}
 \acro{PPM}{piecewise parabolic method}
 \acro{UV}{ultraviolet}
 \acro{HST}{Hubble Space Telescope}
 \acro{COS}{Cosmic Origins Spectrograph}
 \acro{QSO}{Quasar}
 \acro{THERMAL}{Thermal History and Evolution in Reionization Models of Absorption Lines}
 \acro{FDM}{Fuzzy Dark Matter}
 \acro{WDM}{Warm Dark Matter}
 \acro{TDR}{temperature-density relation}
 \acro{GP}{Gaussian process}
 \acro{SED}{spectral energy distribution}
\end{acronym}

\clearpage

\end{document}

%% file: datatable.tex
\begin{deluxetable*}{ccccccc}
\tabletypesize{\footnotesize}
\tablecolumns{7}
\tablewidth{0pt}
\tablecaption{Different data sets used in this analysis}
\tablehead{
 \colhead{dataset} & 
 \colhead{\(z_\mathrm{min}\)}&
 \colhead{\(z_\mathrm{max}\)}&
 \colhead{\(\Delta z\)}&
 \colhead{\(N_\mathrm{qso}\)} &
 \colhead{\(\sim R\)} &
 \colhead{\(k_\mathrm{max} \mathrm{[s/km]}\)}
 }
\startdata
  \label{tab:datasets}
\citet{palanque-delabrouille2013onedimensionalLy} &2.2&4.2&0.2& 11000 & 2200  & 0.02  \\
\citet{viel2013Warmdarkmatter} &4.2&5.4&0.4& 15 & 60000   & 0.1\\
\citet{Irsic2017Newconstraints} &3.0&4.2&0.2& 100 & 6000--9000   & 0.05 \\
\citet{Walther2017NewPrecision} &1.8&3.4&0.2& 74 & 60000   &0.1\\
\enddata
\end{deluxetable*}

%% file: modeltable.tex
\begin{deluxetable}{ccccc}
\tabletypesize{\footnotesize}
\tablecolumns{5}
\tablewidth{0pt}
\tablecaption{Thermal evolution models used in comparisons to existing measurements, parameters are the reionization redshifts and the total heat input during reionization for \HI{} and \HeII{}, see \citet{Onorbe2017Self-consistentModeling} for details}
\tablehead{
 \colhead{model name} & 
 \colhead{\(z_\mathrm{reion,\mathHI}\)}&
 \colhead{\(z_\mathrm{reion,\mathHeII}\)}&
 \colhead{\(\Delta T_{\mathHI}[\si{\kelvin}]\)}& 
 \colhead{{\(\Delta T_{\mathHeII}[\si{\kelvin}]\)}}
 }
\startdata
  \label{tab:thermalpar_modelcomp}
no \HeII &7.3& -- &20000& -- \\
cold \HeII &6.55&3.0&20000&10000\\
standard \HeII &6.55&3.0&20000&15000\\
warm \HeII &6.55&3.0&20000&20000\\
hot \HeII &6.55&3.0&20000&30000\\
late \HeII &6.55&2.8&20000&15000
\enddata
\end{deluxetable}

%% file: MCMC_results_flat.tex
\begin{deluxetable*}{cccccccS}
\tabletypesize{\footnotesize}
\tablecolumns{8}
\tablewidth{0pt}
\tablecaption{Fiducial Evolution of Thermal Parameters Assuming a Flat Prior on \(\bar{F}\). Columns are: redshift \(z\), pressure smoothing scale \(\lambda_\mathrm{P}\), temperature at mean density \(T_0\), \ac{TDR} index \(\gamma\), mean transmission \(\bar{F}\), temperature at optimal overdensity for this measurement \(T_{\Delta_\mathrm{power}}\), temperature at optimal overdensity for curvature analyses \(T…{\Delta_{\star}}\), and the optimal overdensity \(\Delta_{\mathrm{power}}\) for this measurement. Note that quoted error values are purely statistical and  systematic uncertainties of parameters due to box size, fixed cosmology and cosmic variance are expected to be about \(0.5\)--\(1\sigma_\mathrm{stat}\) based on the analysis presented in \autoref{sec:discussion}. However, this estimate is based on a single simulation and a precise characterization of systematics requires a large set of additional simulations. }
\tablehead{
 \colhead{\(z\)}&
 \colhead{\(\lambda_P\)}&
 \colhead{\(T_0\)}&
 \colhead{\(\gamma\)} &
 \colhead{\(\bar{F}\)} &
 \colhead{\(T_{\Delta_{\mathrm{power}}}\)} &
 \colhead{\(T_{\Delta_{\star}}\)} &
 \colhead{\(\Delta_{\mathrm{power}}\)} \vspace{-2.5mm}\\
 \colhead{}&
 \colhead{[\(\si{\kpc}\)]}&
 \colhead{[\(10^4\,\si{\kelvin}\)]}&
 \colhead{} &
 \colhead{} &
 \colhead{[\(10^4\,\si{\kelvin}\)]} &
 \colhead{[\(10^4\,\si{\kelvin}\)]} &
 \colhead{}
 }
\startdata
  \label{tab:results_flat}\vspace{-3mm}\\
1.8 & $79.0_{-11.9}^{+16.0}$ & $0.684_{-0.12}^{+0.18}$ & $1.97_{-0.26}^{+0.16}$ & $0.872_{-0.018}^{+0.020}$ & $1.160_{-0.24}^{+0.24}$ & $4.288_{-1.53}^{+2.16}$ & 1.7047125096061697\\
2.0 & $93.0_{-17.4}^{+8.3}$ & $0.734_{-0.071}^{+0.093}$ & $2.15_{-0.26}^{+0.09}$ & $0.831_{-0.011}^{+0.033}$ & $1.096_{-0.12}^{+0.12}$ & $5.749_{-2.29}^{+1.01}$  & 1.4373354800634492\\
2.2 & $91.0_{-6.4}^{+6.3}$ & $0.789_{-0.068}^{+0.085}$ & $2.13_{-0.13}^{+0.09}$ & $0.796_{-0.009}^{+0.010}$ & $1.369_{-0.11}^{+0.12}$ & $4.942_{-0.77}^{+0.77}$   & 1.637989311491324\\
2.4 & $87.2_{-5.1}^{+5.4}$ & $0.831_{-0.078}^{+0.11}$ & $2.07_{-0.18}^{+0.13}$ & $0.772_{-0.012}^{+0.013}$ & $1.593_{-0.12}^{+0.14}$ & $3.995_{-0.63}^{+0.72}$   & 1.8410421900902019\\
2.6 & $88.3_{-4.5}^{+3.7}$ & $1.000_{-0.090}^{+0.15}$ & $1.93_{-0.17}^{+0.15}$ & $0.745_{-0.013}^{+0.012}$ & $1.936_{-0.084}^{+0.095}$ & $3.449_{-0.35}^{+0.45}$   & 2.0122641166841055\\
2.8 & $93.8_{-4.2}^{+4.2}$ & $1.000_{-0.087}^{+0.11}$ & $2.16_{-0.13}^{+0.09}$ & $0.688_{-0.010}^{+0.013}$ & $1.982_{-0.15}^{+0.16}$ & $3.911_{-0.41}^{+0.43}$   & 1.8180140009876085\\
3.0 & $80.6_{-5.6}^{+6.0}$ & $1.429_{-0.27}^{+0.31}$ & $1.47_{-0.24}^{+0.26}$ & $0.694_{-0.015}^{+0.009}$ & $2.027_{-0.14}^{+0.16}$ & $2.347_{-0.23}^{+0.27}$   & 2.1095466930988134\\
3.2 & $84.9_{-6.3}^{+4.7}$ & $1.115_{-0.15}^{+0.23}$ & $1.85_{-0.25}^{+0.21}$ & $0.623_{-0.019}^{+0.018}$ & $1.910_{-0.15}^{+0.17}$ & $2.465_{-0.25}^{+0.28}$   & 1.8820714405050851\\
3.4 & $90.1_{-5.8}^{+4.7}$ & $1.330_{-0.22}^{+0.30}$ & $1.82_{-0.27}^{+0.24}$ & $0.569_{-0.023}^{+0.021}$ & $2.202_{-0.21}^{+0.21}$ & $2.592_{-0.28}^{+0.28}$   & 1.791247343796632\\
3.6 & $79.4_{-9.6}^{+10.0}$ & $1.010_{-0.30}^{+0.36}$ & $1.74_{-0.36}^{+0.28}$ & $0.512_{-0.021}^{+0.022}$ & $1.160_{-0.34}^{+0.39}$ & $1.704_{-0.56}^{+0.60}$  & 1.1943343873227585\\
3.8 & $79.4_{-6.7}^{+8.4}$ & $1.029_{-0.25}^{+0.29}$ & $1.74_{-0.39}^{+0.29}$ & $0.433_{-0.026}^{+0.025}$ & $1.320_{-0.27}^{+0.36}$ & $1.548_{-0.34}^{+0.44}$   & 1.4422680097183826\\
4.0 & $72.3_{-5.8}^{+7.6}$ & $0.863_{-0.19}^{+0.27}$ & $1.42_{-0.34}^{+0.37}$ & $0.387_{-0.022}^{+0.017}$ & $0.942_{-0.20}^{+0.29}$ & $1.090_{-0.26}^{+0.34}$   & 1.2182648314162368\\
4.2 & $77.0_{-6.0}^{+3.6}$ & $0.905_{-0.082}^{+0.12}$ & $1.73_{-0.40}^{+0.33}$ & $0.355_{-0.031}^{+0.025}$ & $1.051_{-0.082}^{+0.087}$ & $1.246_{-0.16}^{+0.12}$   & 1.2146075359311088\\
4.6 & $73.7_{-5.8}^{+4.9}$ & $0.910_{-0.12}^{+0.12}$ & $1.54_{-0.39}^{+0.37}$ & $0.278_{-0.028}^{+0.023}$ & $0.966_{-0.11}^{+0.13}$ & $1.037_{-0.12}^{+0.15}$   & 1.1344777498695007\\
5.0 & $57.3_{-4.3}^{+4.0}$ & $0.535_{-0.092}^{+0.12}$ & $1.54_{-0.33}^{+0.31}$ & $0.159_{-0.020}^{+0.018}$ & $0.555_{-0.095}^{+0.12}$ & $0.580_{-0.10}^{+0.12}$   & 1.0672879468212193\\
5.4 & $54.4_{-4.5}^{+4.3}$ & $0.597_{-0.13}^{+0.15}$ & $1.55_{-0.29}^{+0.29}$ & $0.060_{-0.008}^{+0.009}$ & $0.551_{-0.12}^{+0.14}$ & $0.613_{-0.14}^{+0.16}$   & 0.8679889292017176\\
\enddata
\end{deluxetable*}

%% file: MCMC_results_strong.tex
\begin{deluxetable*}{cccccccS}
\tabletypesize{\footnotesize}
\tablecolumns{8}
\tablewidth{0pt}
\tablecaption{ Fiducial Evolution of Thermal Parameters Assuming the Strong Prior on \(\bar{F}\). Columns are defined in \autoref{tab:results_flat}. Note that quoted error values are purely statistical and  systematic uncertainties of parameters due to box size, fixed cosmology and cosmic variance are expected to be about \(0.5\)--\(1\sigma_\mathrm{stat}\) based on the analysis presented in \autoref{sec:discussion}. However, this estimate is based on a single simulation and a precise characterization of systematics requires a large set of additional simulations.}
\tablehead{
 \colhead{\(z\)}&
 \colhead{\(\lambda_P\)}&
 \colhead{\(T_0\)}&
 \colhead{\(\gamma\)} &
 \colhead{\(\bar{F}\)} &
 \colhead{\(T_{\Delta_{\mathrm{power}}}\)} &
 \colhead{\(T_{\Delta_{\star}}\)} &
 \colhead{\(\Delta_{\mathrm{power}}\)} \vspace{-2.5mm}\\
 \colhead{}&
 \colhead{[\(\si{\kpc}\)]}&
 \colhead{[\(10^4\,\si{\kelvin}\)]}&
 \colhead{} &
 \colhead{} &
 \colhead{[\(10^4\,\si{\kelvin}\)]} &
 \colhead{[\(10^4\,\si{\kelvin}\)]} &
 \colhead{}
 }
\startdata
  \label{tab:results_strong}\vspace{-3mm}\\
1.8 & $65.9_{-4.2}^{+5.0}$ & $0.768_{-0.22}^{+0.37}$ & $1.63_{-0.25}^{+0.16}$ & $0.897_{-0.005}^{+0.005}$ & $2.011_{-0.28}^{+0.31}$ & $2.533_{-0.38}^{+0.44}$  & 4.760023230765577\\%
2.0 & $75.5_{-6.4}^{+9.8}$ & $0.732_{-0.091}^{+0.17}$ & $1.88_{-0.27}^{+0.20}$ & $0.865_{-0.019}^{+0.015}$ & $1.357_{-0.15}^{+0.20}$ & $3.411_{-0.83}^{+1.32}$  & 1.9830013799905144\\
2.2 & $79.4_{-5.0}^{+5.1}$ & $1.014_{-0.15}^{+0.25}$ & $1.74_{-0.21}^{+0.15}$ & $0.825_{-0.008}^{+0.009}$ & $2.119_{-0.15}^{+0.18}$ & $3.338_{-0.44}^{+0.49}$  & 2.712582375790601\\
2.4 & $81.1_{-4.7}^{+4.6}$ & $1.165_{-0.19}^{+0.29}$ & $1.63_{-0.19}^{+0.16}$ & $0.799_{-0.008}^{+0.008}$ & $2.267_{-0.17}^{+0.19}$ & $2.980_{-0.30}^{+0.35}$  & 2.8279839802572035\\
2.6 & $84.9_{-4.8}^{+4.4}$ & $1.234_{-0.14}^{+0.19}$ & $1.67_{-0.15}^{+0.13}$ & $0.763_{-0.007}^{+0.007}$ & $2.277_{-0.092}^{+0.097}$ & $2.994_{-0.21}^{+0.23}$  & 2.500737683674224\\
2.8 & $91.3_{-5.3}^{+4.5}$ & $1.286_{-0.15}^{+0.19}$ & $1.78_{-0.12}^{+0.11}$ & $0.719_{-0.008}^{+0.008}$ & $2.610_{-0.20}^{+0.22}$ & $3.278_{-0.27}^{+0.30}$  & 2.461799536183449\\
3.0 & $81.7_{-5.9}^{+5.8}$ & $1.289_{-0.14}^{+0.18}$ & $1.60_{-0.16}^{+0.14}$ & $0.687_{-0.008}^{+0.008}$ & $1.946_{-0.14}^{+0.15}$ & $2.408_{-0.21}^{+0.24}$  & 2.015895330177234\\
3.2 & $83.4_{-5.3}^{+5.6}$ & $1.186_{-0.12}^{+0.13}$ & $1.75_{-0.13}^{+0.11}$ & $0.631_{-0.008}^{+0.007}$ & $1.770_{-0.14}^{+0.15}$ & $2.385_{-0.22}^{+0.24}$  & 1.7351817072863045\\
3.4 & $88.7_{-5.3}^{+5.2}$ & $1.404_{-0.16}^{+0.17}$ & $1.74_{-0.11}^{+0.10}$ & $0.576_{-0.007}^{+0.007}$ & $2.075_{-0.21}^{+0.21}$ & $2.555_{-0.27}^{+0.27}$  & 1.634058369851635\\
3.6 & $79.7_{-10.7}^{+9.5}$ & $1.038_{-0.27}^{+0.31}$ & $1.69_{-0.25}^{+0.14}$ & $0.518_{-0.007}^{+0.007}$ & $0.666_{-0.14}^{+0.16}$ & $1.696_{-0.61}^{+0.64}$ & 0.5093612235323396\\
3.8 & $77.8_{-6.9}^{+8.3}$ & $1.205_{-0.19}^{+0.23}$ & $1.41_{-0.23}^{+0.20}$ & $0.457_{-0.006}^{+0.006}$ & $1.132_{-0.18}^{+0.20}$ & $1.524_{-0.33}^{+0.43}$  & 0.8480161843108038\\
4.0 & $71.5_{-5.2}^{+7.4}$ & $0.940_{-0.17}^{+0.22}$ & $1.27_{-0.24}^{+0.24}$ & $0.397_{-0.006}^{+0.006}$ & $0.878_{-0.15}^{+0.19}$ & $1.084_{-0.26}^{+0.33}$  & 0.781583628871446\\
4.2 & $77.5_{-5.4}^{+3.3}$ & $0.890_{-0.073}^{+0.093}$ & $1.85_{-0.33}^{+0.23}$ & $0.346_{-0.022}^{+0.025}$ & $1.047_{-0.079}^{+0.082}$ & $1.268_{-0.15}^{+0.11}$  & 1.2092721197295848\\
4.6 & $76.2_{-5.4}^{+4.2}$ & $0.877_{-0.11}^{+0.13}$ & $1.84_{-0.33}^{+0.23}$ & $0.254_{-0.020}^{+0.021}$ & $1.016_{-0.11}^{+0.14}$ & $1.080_{-0.12}^{+0.15}$  & 1.2026023829194825\\
5.0 & $57.7_{-4.3}^{+4.2}$ & $0.533_{-0.091}^{+0.12}$ & $1.64_{-0.32}^{+0.26}$ & $0.152_{-0.017}^{+0.016}$ & $0.576_{-0.099}^{+0.12}$ & $0.586_{-0.10}^{+0.12}$  & 1.1252143106586299\\
5.4 & $54.3_{-4.6}^{+4.3}$ & $0.599_{-0.13}^{+0.15}$ & $1.54_{-0.29}^{+0.29}$ & $0.061_{-0.008}^{+0.009}$ & $0.549_{-0.12}^{+0.14}$ & $0.616_{-0.14}^{+0.16}$  & 0.8567062758621838\\
\enddata
\end{deluxetable*}